\documentclass[journal,10pt,twocolumn]{IEEEtran}
\usepackage{graphicx, times, amsmath, amsfonts, comment,amsthm}
\usepackage{amssymb, enumitem, epstopdf, soul, color, cite}
\usepackage[noend]{algorithmic}
\usepackage{algorithm, array}

\newcommand{\beq}{\begin{equation}}
\newcommand{\eeq}{\end{equation}}
\newcommand{\tbf}{\textbf}
\newcommand{\tit}{\textit}

\newcommand{\ud}{\mathrm{d}}

\theoremstyle{plain}
\newtheorem{defcounter}{Definition}
\newtheorem{definition}[defcounter]{Definition}
\theoremstyle{plain}
\newtheorem{assumecounter}{Assumption}
\newtheorem{assumption}[assumecounter]{Assumption}
\theoremstyle{plain}
\newtheorem{lemmacounter}{Lemma}
\newtheorem{lemma}[lemmacounter]{Lemma}
\theoremstyle{plain}
\newtheorem{propcounter}{Proposition}
\newtheorem{proposition}[propcounter]{Proposition}
\theoremstyle{plain}
\newtheorem{theoremcounter}{Theorem}
\newtheorem{theorem}[theoremcounter]{Theorem}
\theoremstyle{plain}
\newtheorem{corocounter}{Corollary}
\newtheorem{corollary}[corocounter]{Corollary}
\theoremstyle{plain}
\newtheorem{condcounter}{Condition}
\newtheorem{condition}[condcounter]{Condition}

\newcommand {\Ebb}{\mathbb{E}}

\newcommand {\Rbb}{\mathbb{R}}

\newcommand {\Kcal}{\mathcal{K}}

\newcommand {\Mcal}{\mathcal{M}}

\begin{document}

\title{Self-Sustainability of Energy Harvesting Systems: Concept, Analysis, and Design}
\author{\IEEEauthorblockN{Sudarshan Guruacharya and Ekram Hossain \\ \thanks{The work was supported by a CRD grant (CRDPJ 461412-13) from the Natural Sciences and Engineering Research Council of Canada (NSERC).}} 
\IEEEauthorblockA{Department of Electrical and Computer Engineering, University of Manitoba, Canada \\
Emails: \{sudarshan.guruacharya,ekram.hossain\}@umanitoba.ca}}
\maketitle

\begin{abstract}
Ambient energy harvesting is touted as a low cost solution to prolong the life of low-powered devices, reduce the carbon footprint, and make the system self-sustainable. Most research to date have focused either on the physical aspects of energy conversion process or on the optimal consumption policies of the harvested energy at the system level. However, although intuitively understood, to the best of our knowledge, the idea of self-sustainability is yet to be systematically studied as a performance metric. In this paper, we provide  a mathematical definition of the concept of self-sustainability of an energy harvesting system, based on the complementary idea of \tit{eventual} energy outage, rather than the usual energy outage. In particular, we  analyze a harvest-store-consume system with infinite battery capacity, stochastic energy arrivals, and fixed energy consumption rate. Using the random walk theory, we identify the necessary condition for the system to be self-sustainable. General formulas are given for the self-sustainability probability in the form of integral equations. Since these integral equations are difficult to solve analytically, an exponential upper bound for eventual energy outage is given using martingales. This bound guarantees that the eventual energy outage can be made arbitrarily small simply by increasing the initial battery energy. We then give an  asymptotic formula; and for the special case when the energy arrival follows a Poisson process, we are also able to find an exact formula for the eventual energy outage probability. We also show that the harvest-store-consume system is mathematically equivalent to a $GI/G/1$ queueing system, which allows us to easily find the energy outage probability, in case the necessary condition for self-sustainability is violated. Monte-Carlo simulations  verify our analysis.
\end{abstract}


\section{Introduction}
In recent years, ambient energy harvesting and its applications have become a topic to great interest. Basically, a device is assumed to have an ability to harvest energy from random energy sources in their environment.  Such methods are useful when devices are low powered and energy constrained. They can help to extend the life time of a device, make them self-sustaining, and lower their maintenance cost. Furthermore, they can help lower carbon dioxide emissions and help fight climate change. The readers are referred to \cite{Raghunathan2006a,Chalasani2008,El-Sayed2016,Prasad2014,Xiao2015} for general surveys of this field.

Apart from traditional sources of ambient energy such as sun, wind, and wave, in the past decade research has been extended to energy scavenging techniques from diverse energy sources  \cite{Paradiso2005,Tabbakh2010} such as thermal \cite{Kim2010, Alhawari2013},  pressure and vibrations \cite{Kim2010, Koul2015}, ambient radio-frequency (RF) radiation \cite{Visser2008,Hemour2014}, bodily motions \cite{Magno2016}, magnetic field \cite{Roscoe2013},  ambient sound \cite{Tan2013}, and ambient light \cite{Tan2011,Teran2014}.
Such energy harvesting techniques have been applied in the context of various technologies such as wireless sensor networks \cite{Raghunathan2006b, Seah2009, Stojcev2009, Abdelaal2014}, telecommunications \cite{He2015,Ulukus2015,Ku2016,Zhang2015}, cellular networks \cite{Huang2015a}, cognitive radio network \cite{Huang2015b,Mohjazi2015,Hoang2015}, vehicular network \cite{Atallah2016}, health care \cite{Dangi2014, Kanan2016}, IoT (Internet of things) \cite{Roselli2015,Kamalinejad2015}, IoE (Internet of energy) \cite{Wang2016}, and smart grid \cite{He2016,Hosseinimehr2016,Han2015} technology. 

\subsection{Nature of Energy Sources and General Architectures}
The ambient energy sources can be loosely classified as either \tit{steady energy source} or \tit{intermittent (or bursty) energy source}. These sources can be further classified as \tit{predictable}, \tit{semi-predictable}, or \tit{unpredictable} \cite{Ku2016}.  These diverse sources of energy are converted into electricity, which is then either directly consumed, or stored in a rechargeable  battery or capacitor\footnote{By slight abuse of word, in the paper ``battery'' will refer to any energy storing device.} for future use. As with harvesting, consumption of energy can also be either steady or intermittent. Three general architectures of the energy harvesting systems are \tit{harvest-consume}, \tit{harvest-store-consume}, and \tit{harvest-store/consume}~\cite{Xiao2015,Ku2016}. 

In the \tit{harvest-consume} (HC) model, the harvested energy is immediately consumed by the consumer. This model is appropriate when a steady supply of harvested energy can be guaranteed and a battery-free circuit is desired. The major problem with this approach is that, due to possible random nature of the energy source, when the harvested energy is less than the minimum operational energy required by the consuming device, the device is disabled. We say that the consumer has experienced an \tit{energy outage}. 

In \tit{harvest-store-consume} (HSC) model, the harvested energy is first transferred to a rechargeable battery. The consumer then accesses the harvested energy from the battery. By this method the consumer can ensure a steady supply of energy, even though the harvested energy may be intermittent and randomly varying; thus decoupling the harvest and  the consumption processes. 

Since a battery cannot simultaneously charge and discharge, a more practical HSC model is given by the hybrid \tit{harvest-store/consume} (HS/C). In this model, the harvested energy is directly consumed after being stored in a capacitor. If the harvested energy is above the capacity of the capacitor, then the excess energy is stored in a battery. If the capacitor is empty, then the energy is consumed from the battery. Thus, the HS/C requires dual storage units.  

\subsection{Energy Outage}
A lot of research attention has been paid either on the physical aspects of energy harvesting mechanism \cite{Paradiso2005}-\cite{Teran2014} or on the consumption policies of the harvested energy \cite{Raghunathan2006b}-\cite{Han2015}. Work dealing with former issue tend to focus on physical modeling and optimization of energy harvesting devices, with the goal of improving the efficiency of energy conversion process. Work dealing with the latter issue tend to focus on optimizing some performance metric, under energy constraints of the energy harvesting device. Clearly, the work of the latter category is affected by the general  architecture assumed. Also, the exact nature of the metric depends on what the harvested energy is being used for. For instance, in the field of energy harvesting communication, when the HC model is used, a possible natural performance metric is the joint energy and information outage probability \cite{Flint2015,Xiao2016}. If instead the HSC model is used, a possible performance metric is the data rate of the communication system \cite{Ulukus2015}. 

While there are plenty of work that have examined systems that have energy harvesting capability, very few work have focused on examining the concept of \tit{energy outage} of the consumer, which should be one of the key performance metrics of any energy harvesting system. The consumer is said to experience \tit{energy outage} if there is no energy available for its consumption. For the HC model, the energy outage is determined by the randomness of the harvester-to-consumer energy transmission channel or the inherent randomness of the energy source. In \cite{Flint2015,Xiao2016} the harvested energy is assumed to be a stationary, ergodic process. As such, knowing the distribution of the harvested energy allows us to calculate the energy outage probability. For the HSC model, the harvested energy is often modeled as discrete packets of possibly variable size \cite{Ulukus2015}. When the time is slotted, the energy arrival is usually modeled as a Bernoulli process; and in \cite{Huang2013} the complement of energy outage probability, referred to as ``transmission probability,'' is obtained. The energy arrival process at the battery is generally modeled as a Poisson process, when time is assumed to be continuous, allowing the battery state to be modeled as a Markov chain \cite{Li2015}. However, research on energy outage is sorely lacking for HSC architecture.

\subsection{Self-sustainability and Eventual Energy Outage}
Another concept closely associated with energy harvesting systems is the concept of \tit{self-sustainability}. The main objective of energy harvesting is to make the system self-sustainable. For instance, in \cite{Ulukus2015} the authors state that energy harvesting capability will provide ``self-sustainability and virtually perpetual operation.'' Self-sustainability is commonly understood as 1) the ability to supply one's own needs without external assistance, or 2) the ability to maintain oneself once commenced \cite{Merriam-Webster-2017}.  In the context of energy harvesting systems, the term is used to denote the ability of the system to provide the necessary energy for the consumer, without the need for external power. However, apart from invoking this vague literal meaning, we are not aware of any prior systematic attempt to study this notion from a system theoretical point of view.

A possible quantification of the concept of self-sustainability is to define it as the complement of energy outage probability. This gives us the percentage of time that a system will not need to depend on grid electricity; and it roughly captures the first possible meaning of this word. This is essentially how the term is used in \cite{Xiao2016}. In \cite{Maso2014}, self-sustainability is defined as the ratio of harvested power to the consumed power. We will later show this ratio to be equivalent to the complement of energy outage probability in \tbf{Section~\ref{sec:battery-energy-evolution}}. In this paper, we refer to this ratio as the \tit{utilization factor} instead. However, this definition does not capture the notion of perpetual operation of the system once commenced, as given by the second possible meaning of the word, and indeed as intended by \cite{Ulukus2015}. Thus, our main contribution is to mathematically quantify this vague, intuitive notion of self-sustainability as the perpetual operability of the system and to show how these two possible definitions are related to each other. To the best of our knowledge, this aspect of an energy harvesting system has not been explored. 

In this paper, we define the concept of self-sustainability as follows: \tbf{\tit{The self-sustainability of an energy harvesting system is the probability that the consumer will not eventually experience an energy outage}.} Here we need to make a distinction between \tit{energy outage} and \tit{eventual energy outage}. We say that a consumer experiences an {\em eventual energy outage} if the consumer undergoes an energy outage within \tit{finite time}. If the consumer has to wait for infinite amount of time to experience an energy outage, then we say that the energy harvesting system is \tit{self-sustainable}. The eventual energy outage and the self-sustainability are complementary in that the sum of their probabilities is unity. In other words, self-sustainability of an energy harvesting system refers to the \tit{perpetual, uninterrupted operability of the system}. This is a useful quality to possess when the system is required to have ultra high reliability, especially in mission-critical applications like health or hazard monitoring, where an energy outage of the system can lead to fatal consequences or severe property damages. 

Symbolically, we will denote the {\em self-sustainability probability} by $\phi$ and {\em eventual energy outage probability} by $\psi$ such that $\phi + \psi = 1$. We will say that a system is \tit{self-sustainable} if $\phi > 0$ (or equivalently $\psi < 1$) and unsustainable if $\phi = 0$ (or equivalently $\psi = 1$).

Also, let the energy outage probability be denoted by $P_{out}$. We will see later in \tbf{Section~\ref{sec:battery-energy-evolution}} that $\psi = 1 \Leftrightarrow P_{out} > 0$ and $\psi < 1 \Leftrightarrow P_{out} = 0$. The energy outage probability is the feature of the system after it achieves steady state (i.e. when the system is stationary and ergodic). Thus it is independent of the initial state of the system. On the other hand, the {\em eventual outage probability} is a feature of the system's transient behavior and is strongly dependent on the initial state of the system. Mathematically, these two behaviors are separated by, what we refer to as, the \tit{self-sustainability condition}. If the self-sustainability condition is satisfied, then the system will not achieve steady state. If the self-sustainability condition is violated, then the system will achieve steady state.

\subsection{Synopsis}
\label{subsec:synopsis}
Given these basic definitions, we can quickly point out that for an HC system, if the  energy source is intermittent and the energy consumption is steady, then the consumer will almost surely experience an eventual outage. Hence, the HC system is not a self-sustaining architecture, which is unlike an HSC or HS/C system. Noting that the HSC and the HS/C systems are mathematically equivalent (although their physical implementations could be different),  we want the answers to the following questions about the HSC system:
\begin{enumerate}
\item Under what condition is self-sustainability possible?
\item Can we come up with a formula or a bound for the probability of self-sustainability?
\item How should we design a system, provided a constraint on the eventual energy outage probability?
\end{enumerate}

In this work, we are concerned with the phenomenological aspects of an HSC system, and not with the aspects of optimization. We have tried to answer the above three questions under a suitable set of assumptions. Specifically, we have studied the HSC architecture for the case where energy arrives impulsively, storage capacity is infinite, and the rate of energy consumption is constant. Given these assumptions, the rest of the paper is occupied with the required theoretical analysis, after which a few simple design principles emerge:

\begin{enumerate}
\item Make sure that the rate of energy consumption is strictly less than the rate of harvest (see \tbf{Proposition~\ref{prop:self-sus}}). 
\item Initialize the battery with sufficient energy (see \tbf{Corollary~\ref{coro:phi-is-proper}}).
\item In case an energy outage occurs, restart the system with the same battery initialization (see \tbf{Proposition~\ref{prop:no-of-outages}}).
\end{enumerate}

Let $\lambda$ be the energy arrival rate, $\bar{X}$ be the average energy of an impulse, $p$ be the fixed consumed power, and $u_0$ be the initial battery energy. Then, the first rule is equivalent to the condition $\lambda \bar{X} > p$.  This is a strict inequality. However, $\lambda \bar{X}$ and $p$ can be as close to each other as we please.  This rule guarantees that $\phi > 0$. 

According to the second rule,  assuming that the first rule is satisfied, larger initial battery energy lowers the eventual energy outage probability; but how  large should the initial battery energy be? To answer this, if we constrain the eventual energy outage probability to be $\psi(u_0) = \epsilon$, where $\epsilon \in (0,1)$ is some given grade-of-service, then the required initial energy is at most $u_0 = \frac{1}{r^*}\log (\frac{1}{\epsilon})$, where $r^*$ is some constant, see \tbf{Corollary~\ref{coro:initial-energy}}. 

As for the third rule, so long as the first rule is satisfied, the system will experience only a finite number of energy outages almost surely. Hence we can keep on restarting the system until it ultimately becomes self-sustaining. If we are willing to re-start the system at most $k$ times with the same initial conditions once an energy outage occurs, then constraining this probability at $1-\epsilon$, the required $u_0$ is reduced by factor $k+1$, i.e. $u_0 = \frac{1}{(k+1) r^*}\log (\frac{1}{\epsilon})$, see \tbf{Corollary~\ref{coro:initial-energy-with-restarts}}. 

\subsection{Contributions}
The main contribution of this work is the establishment of the concept of self-sustainability of an energy harvesting system. We specifically study the HSC architecture, for the case where energy arrives impulsively, storage capacity is infinite, and the rate of consumption is constant. We analyze the system based on random walks, renewal theory, and martingales. In particular, we are able to adapt many ideas from the ruin theory of actuarial science, making our work cross-disciplinary. This leads to the following findings:

\begin{enumerate}
\item The necessary condition for HSC system to be self-sustaining is simply that the rate of consumption be strictly less than the rate of harvest. That is, $\lambda \bar{X} > p$, where $\lambda$ is the energy arrival rate, $\bar{X}$ is the average energy of an impulse, and $p$ is the fixed consumed power. In this paper, we refer to this condition as the \tit{self-sustainability condition}. 
\item We provide general formulas for the self-sustainability probability of HSC system. In particular, we demonstrate the relationship between the self-sustainability probability and the maximum of the underlying random walk using three different formulas. We show that the eventual energy outage can be made arbitrarily small by simply increasing the initial battery energy. This leads to simple design guidelines, given the eventual energy outage constraint, as discussed in \tbf{Section~ \ref{subsec:synopsis}}. 
\item Since finding the analytical solution to the integral equations is difficult, using the concept of martingales, we provide an exponential upper bound for the eventual energy outage probability for the HSC system, provided that the self-sustainability condition is satisfied. That is, $\psi(u_0) \leq e^{-r^* u_0}$, where $u_0$ is the initial battery energy and $r^*$ is some constant, giving us a simple estimate.
\item Using the renewal type integral equation for the self-sustainability probability,  we provide an asymptotic formula for the eventual energy outage probability based on the key renewal theorem for defective distributions. That is, $\psi(u_0) \sim C e^{-r^* u_0}$, for some constant $C$.
\item For the special case when the arrival of energy packets is modeled as a Poisson process, we give exact formulas for the eventual energy outage probability. That is, $\psi(u_0) = (1 - \frac{r^* p}{\lambda}) e^{-r^* u_0}$.
\item We prove that the HSC system is mathematically equivalent to a $GI/G/1$ queuing system and provide a translation of terms from one system to another. This allows us to import the results from queueing theory when the self-sustainability condition is not satisfied. While the queuing analogy is certainly not new \cite{Ulukus2015}, we  provide a systematic proof of the equivalence and connect it to the idea of energy outage. We believe this to be a powerful problem solving method which allows us to answer many more questions about the energy harvesting system such as the energy outage probability, e.g. $P_{out} = 1 - \frac{\lambda \bar{X}}{p}$.
\end{enumerate}

From here on, without any ambiguity, we will simply refer to the energy outage as the \tit{outage} and the eventual energy outage as the \tit{eventual outage}.

\subsection{Organization}
The rest of the paper is organized as follows: Section~II discusses the system model and assumptions, and also  defines the probability of self-sustainability.  Section~III gives the random walk analysis of the energy surplus process and the self-sustainability probability. Section~IV gives an exponential upper bound on eventual outage probability (which is the complement of self-sustainability probability). Section~V gives an asymptotic approximation of the eventual outage probability and discusses the computation of the adjustment coefficient, while Section~VI studies the special case when the energy arrival is a Poisson process. Section~VII examines the battery energy evolution process, while Section~VIII gives a numerical verification of the obtained formulas. In Section~IX, we discuss a simple application in the context of communication system and future work. Section~X concludes the paper.


\section{System Model, Assumptions, and the Concept of Self-Sustainability}
\subsection{Definitions and Assumptions}
We consider an HSC system in which all the harvested energy is first collected, before being consumed. Thus, the consumer obtains the harvested energy indirectly from the rechargeable battery. The harvested energy can arrive into the battery in a continuous fashion (as in solar or wind) or in an impulsive fashion (as in body motions). We will restrict our analysis to the case of impulsive energy arrivals. Continuous energy arrivals can be converted into impulsive energy arrival by appropriate sampling. Thus, our main physical assumption is as follows:

 \begin{assumption} 
 Harvested energy arrives as impulses into the storage system.
 \end{assumption}

In other words, the harvested energy arrival is a countable process. The harvested energy arrives in the form of packets into the battery, and the size of each energy packet may vary randomly. As such, we can model the \tit{energy surplus process} of the system at any time $t$ as 
\beq
U(t) = u_0 - \int_0^t p(u, t) \ud t + \sum_{i=1}^{N(t)}  h(t - t_i; X_i).
\label{eqn:basic-model}
\eeq
Here $u_0 \geq 0$ is the initial battery energy and  $p(u,t) \geq 0$ is the power consumption from the battery, i.e. $\frac{\ud u}{\ud t} = p(u,t)$, where $u$ is the instantaneous value of $U(t)$. The $X_i \in \Rbb_+$  is the amount of energy in an $i$-th energy packet that arrives at time $t_i$, while  $N(t) = \max\{i : t_i \leq t\}$ is the total number of energy packets that have arrived at the battery by time $t$. Lastly, $h(t; X)$ is the transient of battery charging process given $X$, such that $h(t; X) = 0$ for $t < 0$ and $\lim_{t\to\infty}h(t; X) = X$. 

For example, if energy is delivered to the battery for a short period of time from $t=0$ to $T$ at a fixed rate $q$, then the total energy transferred is $X = qT$. The charging process of the battery is given by a piecewise ramp function 
\[ h(t; X) = \left\{ \begin{array}{lcr} 
				qt,  & \mathrm{if} & 0 \leq t \leq T, \\
				X,   & \mathrm{if} & t > T. 
				\end{array} \right. \]
Re-written, we have $h(t;X) \equiv qT g(t) = X g(t)$, where $g(t)$ is given by a normalized version of the above piecewise ramp function
\[ g(t) = \left\{ \begin{array}{lcr} 
				\frac{t}{T}, & \mathrm{if} & 0 \leq t \leq T, \\
				1,  & \mathrm{if} & t > T. 
				\end{array} \right. \]
Thus, the simplest manner in which $X$ can modify $g(t)$, an underlying transient  function, is by scaling its amplitude, $h(t; X) \equiv X g(t)$. For very short period $T \to 0$, the $g(t)$ can be idealized as a unit step function. We can then neglect the ramp part and simply account for the total energy transferred as $h(t;X) = X$. 

Here, the randomness of $X = qT$ can be due to randomness of $q$ or $T$ or both. In the study of wireless power transfer, it is generally assumed that $q$ randomly varies from one impulse to another due to multi-path fading, but is fixed over the short duration $T$. Here $T$ is constant, and this makes $g(t)$ deterministic. However, if the randomness arises due to variable $T$, then $g(t)$ itself is a random function. We can express $h(t; X)$ given $X$ as $h(t;X) = X g(t;X)$ where
\[ g(t; X) = \left\{ \begin{array}{lcr} 
				\frac{qt}{X}, & \mathrm{if} & 0 \leq t \leq X/q, \\
				1, & \mathrm{if} & t > X/q. 
				\end{array} \right. \]				

Clearly, if $U(t) > 0$, then the system is producing more energy than it is consuming. Likewise, if $U(t) < 0$, then the consumer takes in energy from the grid to compensate for energy deficit. We will now define a few concepts that will be used in the paper:

\begin{definition}[Defective and proper distributions]
A random variable with distribution $F$ is said to be \tbf{defective} if $\lim_{x\rightarrow\infty}F(x) < 1$, the amount of defect being $1 - F(\infty)$. The random variable is said to be \tbf{proper} if $\lim_{x\rightarrow\infty}F(x) = 1$.
\label{def:proper-defective-rv}
\end{definition}

\begin{definition}[Renewal process]
A sequence $\{S_n\}$ is a \tbf{renewal process} if $S_n = X_1 + \cdots + X_n$ and $S_0 = 0$, where $\{X_i\}$ are mutually independent, non-negative random variables with common distribution $F_X$ such that $F_X(0) = 0$. When $F_X$ is a proper distribution, the renewal process is said to be a \tbf{persistent} renewal process. If $F_X$ is defective, then the renewal process is said to be a \tbf{transient} or \tbf{terminating} renewal process.
\label{def:renewal-process}
\end{definition}

The variable $X_i$ in \tbf{Definition \ref{def:renewal-process}} is often interpreted as ``inter-arrival time''. However, $X_i$ does not need to be always time, as we will see later in the paper. We will now make further mathematical assumptions required to simplify our analysis.  

\begin{assumption}
The storage capacity of the battery is infinite.
\end{assumption}

\begin{assumption}
The rate of energy consumption is constant, i.e. $p(u,t) = p$.
\end{assumption}

\begin{assumption}
The battery charging process is $h(t; X) = X g(t)$, where $g(t)$ is a unit step function.
\end{assumption}

\begin{assumption}
The energy packets $\{E_i\}$ arrive at corresponding times $\{t_i\}$, for $i = 0, 1, 2 \ldots,$ such that $0 = t_0 < t_1 < t_2 < \cdots$, where the energy packet $E_0$ arrives at time $0$. The arrival times $\{t_i\}$ follow (i) a renewal process. That is, the inter-arrival times $\{A_i\}$, where $A_i = t_{i+1} - t_i$, are such that $A_0, A_1, A_2, \ldots$ are independent and identically distributed, with common distribution $F_A$. (ii) To avoid more than one energy arrivals at a time, we assume $F_A(0) = 0$. (iii) Lastly, the expected inter-arrival time is finite, $\Ebb[A_i] < \infty$.
\end{assumption}

\begin{assumption}
The amount of energy $\{X_i\}$ corresponding to the energy packets $\{E_i\}$ for $i=0, 1, 2, \ldots,$ is (i) a non-negative, continuous random variable; (ii) $\{X_i\}$ are independent and identically distributed, with common distribution $F_X$; 
and (iii) lastly, the expected energy in a packet is finite,  $\Ebb[X_i] < \infty$.
\end{assumption}

\begin{assumption}
The inter-arrival times $\{A_i\}$ and the energy packet sizes $\{X_i\}$ are independent of each other.
\end{assumption}

Here \textbf{Assumptions 2--4} essentially simplify the physical setup of our problem, whereas \textbf{Assumptions 5--7} are essentially technical in nature to facilitate the mathematical analysis. Note that in \textbf{Assumption 5}, while not necessary, we have specifically assumed that there is an arrival of an energy packet at time zero. 
In \tbf{Assumption 5}, the renewal process becomes a Poisson process when the inter-arrival times are exponentially distributed. The Poisson process can also arise as a result of superposition of common renewal processes \cite{Cox1954}. Thus, Poisson energy arrival can be used to model situations where multiple harvesters send energy, according to a common renewal process, to a common battery. Likewise, when the inter-arrival time is deterministic, the renewal process can model the slotted time models. Hence, the renewal process is a generalization of these special cases.

\begin{figure}
\begin{center}
	\includegraphics[width=3.75in]{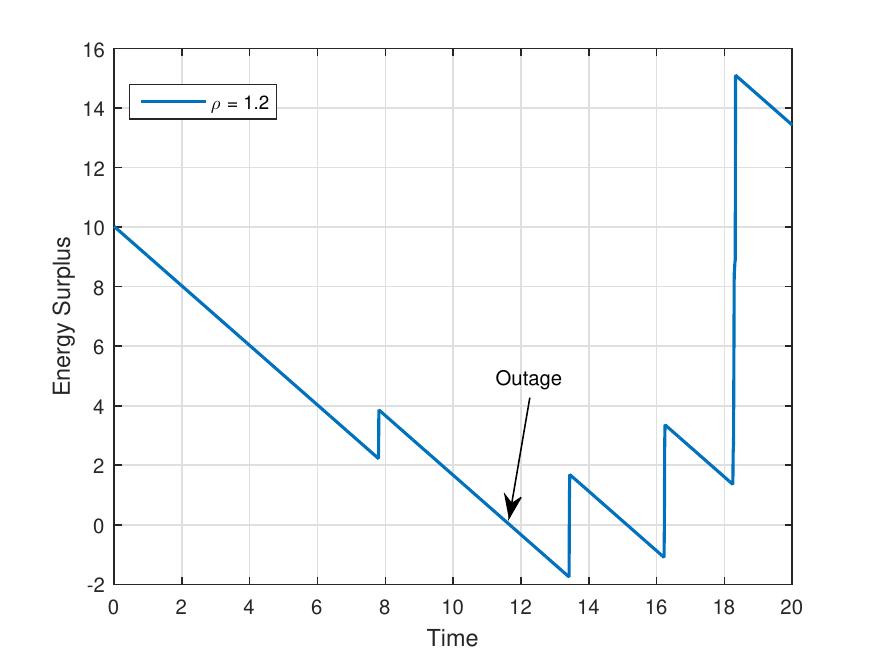}
	\caption{Energy surplus, $U(t)$, versus time, $t$, where $\rho = \frac{\Ebb[X]}{p\Ebb[A]}$.}
	\label{fig:energy-surplus-time}
 \end{center}
\end{figure}

Given these assumptions, our initial mathematical model of the energy surplus (\ref{eqn:basic-model}) simplifies to 
\beq 
U(t) = u_0 - pt + \sum_{i=0}^{N(t)} X_i.
\label{eqn:energy-surplus}
\eeq
Note that battery imperfections like self-discharge can be considered as consumption behavior. Equation (\ref{eqn:energy-surplus}) is an instance of a random walk on real line through continuous time with downward drift. As shown in Fig. \ref{fig:energy-surplus-time}, the graph of $U(t)$ versus $t$ will look like a sawtooth wave with descending ramps, with random jump discontinuities. The system experiences an outage just before $t = 12$.

\subsection{Dual Model}
It is also possible to have a \tit{dual model} of the primal model given in (\ref{eqn:basic-model}). Instead of \tbf{Assumption 1} and \tbf{Assumption 3}, for the dual model we have the following modifications:

\tbf{Assumption 1'.} \tit{The harvested energy arrives at a constant rate.} 

\tbf{Assumption 3'.} \tit{The energy consumed is impulsive.} 

Then, we have the dual model given by
\beq
U(t) = u_0 - \sum_{i=1}^{N(t)} h(t - t_i;  X_i) + \int_0^t p(u, t) \ud t.
\label{eqn:dual-model}
\eeq
Notice that in this dual model there is only a change in the signs and an interchange in the interpretation of the symbols. Thus, the results for our primal model can be shown to be valid for the dual model as well, by trivial variations in the arguments. The dual model partially rectifies the possible criticism regarding constant power consumption in the primal model. From the energy harvesting communication point of view, the dual model is more suitable when there is a steady source of energy (e.g. wind, solar, or RF beacon), while the consumption is discrete stochastic. In the rest of our work, we will focus on the simplified primal model given in (\ref{eqn:energy-surplus}).


\subsection{Concept of Self-Sustainability}
Let $W(t) \geq 0$ be the battery energy at time $t$. We say that the system undergoes \tit{outage} when $W(t) = 0$. In other words, the case $W(t) = 0$ represents the situation when the battery is empty, and as such, the consumer needs to fetch its required energy from the grid to sustain its consumption. We define the outage probability of the system as 
\beq
P_{out} = P(W(t) = 0),
\eeq
that is, probability of finding the battery empty at any time $t$. The $P_{out}$ depends on the steady state behavior of $W(t)$.

Now consider the \tit{first time} that the battery is empty, $\tau = \inf\{t > 0 : W(t) = 0, W(0) = u_0\} = \inf\{t > 0: U(t) \leq 0, U(0) = u_0\}$. We will refer to $\tau$ as the \tit{first time to outage}. If $\tau = \infty$, then the system becomes self-sustaining and the system will not experience an \tit{eventual outage}. The occurrence of eventual outage is equivalent to the event $\{\tau < \infty\}$. However, since $\tau$ itself is a random variable, we can only describe it probabilistically. Thus, we are interested in knowing the probability 
\[ \phi(u_0, p, T) = P\left[\sup_{0\leq t \leq T} \left(pt - \sum_{i=0}^{N(t)} X_i \right) \leq u_0 \right],  \]
that the energy surplus $U(t)$ will not fall below zero through time $t \in [0,T]$ where $T < \infty$, and 
\[ \phi(u_0, p, \infty) = P\left[\sup_{0\leq t \leq \infty} \left(pt - \sum_{i=0}^{N(t)} X_i \right) \leq u_0 \right],  \]
of avoiding an eventual outage. Note that this is equivalent to the probability
\[ \phi(u_0, p, T) = 1 - P(\tau < T| u_0, p) = 1 - \psi(u_0, p, T), \]
where $\psi(u_0,p,T)$ is the probability of an outage occurring within a finite time $T$.

\begin{definition}
We define the \tbf{self-sustainability probability} of an energy harvesting system as the probability that the first time to outage $\tau$ is at infinity. Likewise, the \tbf{eventual outage probability} is the probability that the first time to outage $\tau$ is finite. That is, 
\beq
\phi(u_0, p, \infty) = 1 - P(\tau < \infty|u_0, p) = 1 - \psi(u_0, p, \infty),
\eeq
where $\phi(u_0, p, \infty)$ denotes the self-sustainability probability and $\psi(u_0, p, \infty)$ denotes the eventual outage probability.
\label{def:self-sustainability}
\end{definition}

Since $p$ is held constant and $T = \infty$, from here on, we will drop these two parameters in the argument, and simply refer to the self-sustainability probability and eventual outage probability as a function of $u_0$ as given by $\phi(u_0)$ and $\psi(u_0)$. Thus, $u_0$ is the control parameter.


\section{Energy Surplus Process and Evaluation of Self-Sustainability Probability}

Before we proceed with further analysis, we will give a general proposition on the relationship between the number of outages a system will experience and the self-sustainability probability. 

\begin{proposition}
\label{prop:no-of-outages}
Let the system restart with the same initial battery energy $u_0$ after an outage. Then assuming $\phi>0$,  the number of outages that the system will experience is given by geometric distribution
\beq
P(k \; \mathrm{outages}) = \psi^k \phi, \quad k = 0, 1, 2, \ldots.
\label{eqn:no-of-outage}
\eeq

\end{proposition}
\begin{IEEEproof}
The occurrence of an outage within finite time is a Bernoulli random variable. As such, whenever the system restarts with the same initial setting, we have independent Bernoulli trials. Thus, the probability of having $k$ outages before being outage free is given by geometric distribution. 
\end{IEEEproof}

\begin{corollary}
Assuming $\phi > 0$ and that the system restarts with the same initial configurations after an outage, the system will experience a finite number of outages almost surely.
\end{corollary}
\begin{IEEEproof} 
Since $\phi >0$ and $\psi = 1 - \phi < 1$, we have from (\ref{eqn:no-of-outage}), $P(\mathrm{finite\;outages}) = P(\cup_{k=0}^\infty \{ \mathrm{k\;outages} \}) = \sum_{k=0}^\infty \psi^k \phi = \frac{\phi}{1 - \psi} = 1$.
\end{IEEEproof}

Now the question is,  when will $\phi > 0$? Let the expected inter-arrival time be $\Ebb[A_i]$ and the arrival rate of energy packets be defined as $\lambda = \frac{1}{\Ebb[A_i]}$. Also, let the average energy packet size be $\Ebb[X_i] = \bar{X}$. 

Intuitively, if we want our system to be self-sustainable, then we would want the expected surplus energy to be positive $\Ebb[U(t)] > 0$. For this to be true, it is sufficient that the consumption rate, $p$, be less than the energy arrival rate, $\lambda \bar{X}$. That is, $\lambda \bar{X} > p$. However, the satisfaction of the condition $\lambda \bar{X} > p$ does not mean that an outage will never occur. Rather, it tells us that there is a chance of such non-occurrence of outage.

\subsection{Random Walk Analysis}
In general, it is difficult to characterize the outage event. We therefore have to condition on the arrival process. We can tell that an outage has occurred if a new arrival finds the battery empty. The mathematical trick here is to reduce the continuous time process into a discrete time process by counting over the arrivals. This will allow us to use the techniques from the random walk theory. In the analysis of random walks and ascending ladder process, we will basically follow the approach laid out by Feller in \cite{Feller1971}. A modern treatment of the subject can be found in \cite{Asmussen2003}.

\begin{lemma}
Let the sequences $\{Z_i; i \geq 0\}$ and $\{S_i; i \geq 1\}$ be defined as $Z_i = pA_i - X_i$ and $S_n = \sum_{i=0}^{n-1} Z_i$ where $S_0 = 0$. The surplus energy $\{U_n\}$, observed immediately before the arrival of $n$-th energy packet, forms a discrete time random walk over real line, such that 
\beq 
U_n = u_0 - S_n. 
\label{eqn:random-walk-lemma}
\eeq
\end{lemma}
\begin{IEEEproof}
Let us denote $t_n^- = t_n - \epsilon$ where $\epsilon > 0$, as time immediately before the arrival of $n$-th energy packet at $t_n$. Since $t_n^- \sim t_n $ as $\epsilon \rightarrow 0$, we can decompose the arrival time $t_n^-$ as $t_n^- = \sum_{i=0}^{n-1} A_i$. If we follow the value of $U(t)$ immediately before each arrival at $t_n^-$, we have
\begin{align*}
U_n &= u_0 - p t_n^- + \sum_{i=0}^{N(t_n^-)} X_i \\   
& = u_0 - \sum_{i=0}^{n-1} (p A_i - X_i).
\end{align*}
Let $Z_i = p A_i - X_i$ for $n = 0, 1, 2, \ldots$. Here, $Z_i$ can take any real value, $Z_i \in \Rbb$. The two sequences $\{X_i\}$ and $\{A_i\}$ are independent of each other. Also, the sequence $\{X_i\}$ is independent and identically distributed in itself; and so is $\{A_i\}$. Thus, $\{Z_i\}$ are independent and identical to each other too. Our expression now becomes $U_n = u_0 - \sum_{i=0}^{n-1} Z_i$, which is a discrete time random walk over real line. Defining $\{S_n; n \geq 1\}$ such that $S_n = \sum_{i=0}^{n-1} Z_i$ with initialization $S_0 = 0$, gives us $U_n = u_0 - S_n$, which is our desired result.
\end{IEEEproof}

\tbf{Remark:} Here, the $\{U_i\}$ records the troughs of the sawtooth wave $U(t)$ and can be interpreted as the energy surplus that the $n$-th arrival finds the system in.\footnote{For the dual model, we observe the value of $U(t)$ immediately \tit{after} the departure of an energy packet.} 

The existence of self-sustainable system can now be directly proved with the help of the following theorem from the theory of random walk on real line:

\begin{theorem}\cite[Ch 8, Th 2.4]{Asmussen2003}
For any random walk with $F_Z$ not degenerate at 0, one of the following possibilities occur:
\begin{enumerate}
\item (Oscillating Case) If $\Ebb[Z_i] = 0$, then $P(\lim \sup_{n\rightarrow\infty} \; S_n = +\infty) = 1$,  $P(\lim \inf_{n\rightarrow\infty}\; S_n = -\infty) = 1$; 
\item (Drift to $+\infty$) If $\Ebb[Z_i] > 0$, then $P(\lim_{n\rightarrow\infty} S_n = +\infty) = 1$; 
\item (Drift to $-\infty$) If $\Ebb[Z_i] < 0$, then $P(\lim_{n\rightarrow\infty} S_n = -\infty) = 1$.
\end{enumerate}
\label{theorem:random_walk}
\end{theorem}

The following propositions easily follow from the above theorem.

\begin{proposition}
The HSC system is self-sustainable only if it satisfies the self-sustainability condition given by $\lambda \bar{X} > p$. 
\label{prop:self-sus}
\end{proposition}

\begin{IEEEproof}
Consider the contra-positive of the statement: If $\lambda \bar{X} \leq p$, then the HSC system will experience eventual outage. Since $\Ebb[Z_i] = \Ebb[pA_i - X_i] = p \Ebb[A] - \bar{X}$, we have $\lambda \bar{X} \leq p$ to be equivalent to $\Ebb[Z_i] \geq 0$.

If $\lambda \bar{X} < p$, then this is equivalent to the condition $\Ebb[Z_i] > 0$. Hence, from Case~2 of Theorem~\ref{theorem:random_walk}, $S_n \rightarrow +\infty$ almost surely. Since $U_n = u_0 - S_n$ from (\ref{eqn:random-walk-lemma}), by the definition of limit,
\begin{align*}
P(\lim_{n\rightarrow\infty} S_n = +\infty) &= P(\forall c, \; \exists n_0 : \forall n > n_0, \; S_n > c) \\
&= P(\forall c , \; \exists n_0 : \forall n > n_0, \; u_0 - U_n > c) \\ 
&= P(\forall c, \; \exists n_0 : \forall n > n_0, \; U_n < u_0 - c).
\end{align*}

Taking $c = u_0$, we have $P(\exists n_0 : \forall n > n_0, \; U_n < 0) = 1$. That is, $U_n < 0$ almost surely. Thus, the HSC system will experience eventual outage.

Similarly, if $\lambda \bar{X} = p$, this is equivalent to the condition $\Ebb[Z_i] = 0$. Thus, according to the Case~1 of Theorem~\ref{theorem:random_walk} $\lim \sup_{n\rightarrow\infty} \; S_n = +\infty$ and   $\lim \inf_{n\rightarrow\infty}\; S_n = -\infty$ almost surely. It suffices to consider the former case. From the definition of limit supremum, we have
\begin{align*}
& P(\lim\sup_{n\rightarrow\infty}\; S_n = + \infty) =  P(\lim_{n\rightarrow\infty} (\sup_{m \geq n} S_m) = + \infty) \\
= & \; P(\forall c, \; \exists n_0 : \forall n > n_0, \; \sup_{m \geq n} S_m > c) \\
= & \; P(\forall c, \; \exists n_0 : \forall n > n_0, \; \sup_{m \geq n} (u_0 - U_m) > c) \\
= & \; P(\forall c, \; \exists n_0 : \forall n > n_0, \; u_ 0 - \inf_{m \geq n} U_m > c) \\
= & \; P(\forall c, \; \exists n_0 : \forall n > n_0, \;  \inf_{m \geq n} U_m < u_0 - c). 
\end{align*}
Taking $c = u_0$, we have that $P(\exists n_0 : \forall n > n_0, \;  \inf_{m \geq n} U_m < 0) = 1$. Here the $\inf_{m \geq n} U_m < 0$ implies the existence of a subsequence $U_{m_k}$ which is strictly less than zero. 

Thus, combining the cases for $\lambda \bar{X} < p$ and $\lambda \bar{X} = p$, we conclude that the HSC system will experience eventual outage.
\end{IEEEproof}

\tbf{Remark:} Since $\lambda \bar{X} > p$ is a necessary, but not sufficient, condition for self-sustainability, its satisfaction does not guarantee that outage will not occur. However, it provides an easy to check condition under which self-sustainability is possible. Note that strict inequality has to be maintained, even though $p$ can be as close to $\lambda \bar{X}$ as we please. We will refer to this condition $\lambda \bar{X} > p$ (or equivalently $\Ebb[Z_i] < 0$) as the \tbf{\tit{self-sustainability condition}}. 

\begin{condition}[Self-sustainability]
An HSC system is said to be self-sustainable when $\lambda \bar{X} > p$, or equivalently, $\Ebb[Z_i] < 0$.
\end{condition}

A sufficient condition that guarantees self-sustainability is trivially given by $p=0$. This represents the battery recharge process. In this paper, we are exclusively interested in the case when $p>0$, and we implicitly assume this to be the case.

In the following sections, we will investigate two mathematical cases that arise when this condition is or is not satisfied. Below we give an immediate corollary of the \tbf{Proposition~\ref{prop:self-sus}}.

\begin{corollary}
The HSC system will experience an eventual outage almost surely if the self-sustainability condition is not satisfied. If the self-sustainability condition is satisfied, then the probability of eventual outage will be less than unity. That is,
\[ \psi(u_0) \left\{ \begin{array}{lcr}
				= 1, & \mathrm{if} & \lambda \bar{X} \leq p, \\
				< 1, & \mathrm{if} & \lambda \bar{X} > p. 				
				\end{array} \right. \]
\end{corollary}

\subsection{Ascending Ladder Process}
Now that we have succeeded in converting a continuous time process into discrete time process, consider the random walk $\{S_n; n\geq1\}$ as $S_n= Z_0 + \cdots + Z_{n-1}$, or recursively as $S_{n+1} = S_{n} + Z_{n}$, with initial value $S_0 = 0$. Thus, we have $U_n$ as $U_n = u_0 - S_n$. 

Consider the ascending ladder process defined by $M_n = \sup_{0\leq i \leq n} S_i$, which is the partial maximum of partial sums. Since $M_0 = S_0 = 0$, the $\{M_i; i \geq 1\}$ is a positive non-decreasing sequence, hence the name ascending ladder process. We can also relate the values of $M_n$ by the recursion $M_n = \sup(M_{n-1}, S_n)$. Also, let $M = \sup_{i\geq0} S_i$ be the maximum value attained by $S_i$ through the entire duration of its run. Now, the key observation is  that if $M<\infty$, then the eventual outage is equivalent to $\{\tau<\infty\} \equiv \{M>u_0\}$. Hence, we have the following lemma:
\begin{lemma} 
If $M < \infty$, then
\beq 
\phi(u_0) = F_M(u_0) \quad \mathrm{and} \quad \psi(u_0) = 1 - F_M(u_0).  
\label{eqn:sus-outage-in-terms-of-F-M}
\eeq
\end{lemma}
Given the sequence $\{S_i; i \geq 1\}$, the first strict ascending ladder point $(\sigma_1, H_1)$ is the first term in this sequence for which $S_i > 0$. That is, 
$\sigma_1 = \inf\{n \geq 1 : S_n > 0\}$ and $H_1 = S_{\sigma_1}$. In other words, the  epoch of the first entry into the strictly positive half-axis is defined by 
\beq
\{\sigma_1 = n \} = \{ S_1 \leq 0, \ldots, S_{n-1} \leq 0, S_n > 0  \}. 
\eeq
The $\sigma_1$ is called the first \tit{ladder epoch} while $H_1$ is called the first \tit{ladder height}. Let the joint distribution of $(\sigma_1, H_1)$ be denoted by
\beq 
P(\sigma_1 = n, H_1 \leq x) = F_{H,n}(x). 
\label{eqn:joint-dist-ladder-points}
\eeq
The marginal distributions are given by
\begin{align*} 
P(\sigma_1 = n) &= F_{H,n}(\infty), \\
P(H_1 \leq x) &= \sum_{n=1}^\infty F_{H,n}(x) = F_H(x).
\end{align*}
The two variables have the same defect $1 - F_H(\infty) \geq 0$.

We can iteratively define the ladder epochs $\{\sigma_n\}$ and ladder heights $\{H_n \}$ as 
\begin{align*} 
\sigma_{n+1} &= \inf\{ k \geq 1 : S_{k + \xi_n} > S_{\xi_n} \},  \\
H_{n+1} &= S_{\xi_{n+1}} - S_{\xi_n},
\end{align*}
where $\xi_n = \sum_{i=1}^n \sigma_i$. The pairs $(\sigma_i, H_i)$ are mutually independent and have the same common distribution given in (\ref{eqn:joint-dist-ladder-points}). These ladder heights are related to record maximum at time $n$ by $M_{n} = \sum_{i=1}^{\pi_n} H_i$, where $\pi_n$ is the number of ladder points up until time $n$, i.e. $\pi_n = \inf\{k : \sigma_1 + \cdots + \sigma_k \leq n\}$.

An important observation related to the ladder points is that the sums of $\{\sigma_i\}$ and $\{H_i\}$, $\sigma_1 + \cdots + \sigma_n$ and $H_1 + \cdots + H_n$, form (possibly terminating) renewal processes with inter-renewal interval $\sigma_i$ and $H_i$. Clearly, we can have $M < \infty$ if and only if the ascending ladder process is terminating. As per \tbf{Definition \ref{def:renewal-process}}, this renewal process is terminating if the underlying distribution $F_H$ is defective. 

Accordingly, let $H$ be a defective random variable with $F_H(0) = 0$ and $F_H(\infty) = \theta < 1$, then the amount of defect given by $1 - \theta$ represents the probability of termination. In other words, $\theta$ represents the probability of another renewal, while $1-\theta$ represents the probability that the inter-renewal interval is infinite. Thus the termination epoch is a Bernoulli random variable with ``failure'' being interpreted as ``termination''  with probability $1 - \theta$.  The sum $H_1 + \cdots + H_n$ has a defective distribution given by the $n$-fold convolution $F_H^{(n)}$, whose total mass equals \beq 
F_H^{(n)}(\infty) = F^n_H(\infty) = \theta^n.
\label{eqn:total-conv-weight}
\eeq 
This is easily seen by re-interpreting the $n$ ladder epochs as $n$ ``successes'' of a Bernoulli random variable. The defect $1 - \theta^n$ is thus the probability of termination \tit{before} the $n$-th ladder epoch. 

Let us now define the \tit{renewal function} for the ladder heights by the sum on $n$-fold convolutions of $F_H$ as
\beq
\zeta(x) = \sum_{n=0}^\infty F_H^{(n)} (x),
\label{eqn:renewal-function}
\eeq
where $F_H^{(n)}$ is the $n$-fold convolution of $F_H$ defined recursively as $F_H^{(i)}(x) = \int F_H^{(i-1)}(x - t) \ud F_H(t)$ where $i = 1,2, \ldots, n$; and $F_H^{(0)}(x) $ is a unit step function at the origin.

The Laplace transform of $F_H^{(n)}$ can be obtained recursively as: $\widehat{F}_H^{(0)}(r) = 1/ r$, being a unit step function, and so on, until  $\widehat{F}_H^{(n)}(r) = [\widehat{f}_H(r)]^n/r$. Thus, the Laplace transform of the renewal function is \[ \widehat{\zeta}(r) = \frac{1}{r} \sum_{n=0}^\infty [\widehat{f}_H(r)]^n = \frac{1}{r(1 - \widehat{f}_H(r))},\] since for the geometric sum $|\widehat{f}_H(r)| < \widehat{f}_H(0) = 1$ for $r>0$.

The renewal function $\zeta(x)$ is equivalent to the expected number of ladder points in the strip $[0, x]$, where the origin counts as a renewal epoch. Thus, $\zeta(0) = 1$. For a terminating renewal process, the expected number of epochs ever occurring is finite, as from (\ref{eqn:total-conv-weight}), we have
\[ \zeta(\infty) = \sum_{n=0}^\infty F_H^{(n)} (\infty) = \sum_{n=0}^\infty \theta^n =\frac{1}{1 - \theta}. \]
If the ascending ladder process terminates after $n$-th epoch, then $H_1 + \cdots + H_n = M$, the all time maximum attained by the random walk $S_n$. The probability that the $n$-th ladder epoch is the last and that $\{M \leq x\}$ is given by 
\beq 
P(M \leq x, \mathrm{terminate \; after\; } n) = (1 - \theta) F_H^{(n)}(x).
\label{eqn:joint-dist-M-n}
\eeq 
Using (\ref{eqn:total-conv-weight}), the marginalization of (\ref{eqn:joint-dist-M-n}) over $M$ shows us that the probability of the ladder process terminating \tit{after} the $n$-th epoch follows a geometric distribution:
\[ P(\mathrm{terminate \; after \;} n) = (1 - \theta) F_H^{(n)}(\infty) = (1 - \theta) \theta^n. \]
Similarly, marginalizing (\ref{eqn:joint-dist-M-n}) over $n$, we have 
\beq 
P(M \leq x) = (1-\theta) \sum_{n=0}^\infty F_H^{(n)} (x)  = (1 - \theta) \zeta(x). 
\label{eqn:distr-of-M}
\eeq
 
We now need a criteria to determine whether the ascending ladder process terminates or not, as well as a method to find the value of $\theta$. The following proposition also immediately follows from \tbf{Theorem~\ref{theorem:random_walk}} and our discussion about ascending ladder process: 

\begin{proposition}
Given the self-sustainability condition, $\lambda \bar{X} > p$, the ascending ladder height process $\{ H_i \}$ of an HSC system is terminating almost surely. The probability of self-sustainability given in terms of the renewal function $\zeta$ is 
\begin{align}
\phi(u_0) &= (1-\theta) \zeta(u_0).
\label{eqn:selfsus-renewal-fun}
\end{align}
\label{prop:selfsus-renewal-fun}
\end{proposition}
\begin{IEEEproof}
From Case~3 of \textbf{Theorem~\ref{theorem:random_walk}}, since the random walk $S_n$ drifts to $-\infty$ when $\Ebb[Z_i] < 0$, the maximum $M<\infty$ almost surely, and thus $\{H_i\}$ terminates. Equation (\ref{eqn:selfsus-renewal-fun}) is obtained from (\ref{eqn:sus-outage-in-terms-of-F-M}) and (\ref{eqn:distr-of-M}).
\end{IEEEproof}

We now relate the self-sustainability probability with two convolution formulas.

\begin{proposition}
Given the self-sustainability condition, the self-sustainability probability satisfies the following equivalent integral equations:
\begin{align}
\phi(u_0) &= (1 - \theta) + \int_0^{u_0} \phi(u_0 - x) f_H(x) \ud x, 
\label{eqn:renewal-self-sus-1} \\
\phi(u_0) &= \int_0^\infty \phi(x) f_Z(u_0 - x) \ud x.
\label{eqn:renewal-self-sus-2}
\end{align}
\label{prop:outage-convolution-formula}
\end{proposition}
\begin{IEEEproof}
We begin with the fact that $\phi(u_0) = P(M \leq u_0)$ from (\ref{eqn:sus-outage-in-terms-of-F-M}). The proofs of the two equations follow from the standard renewal type argument: 

\vspace{0.1cm}
(1)  The event $\{M \leq u_0\}$ occurs if the ascending ladder process terminates with $M_0$ or else if $H_1$ assumes some positive value $x \leq u_0$ and the residual process attains the age $\leq u_0 - x$. So, 
\begin{align*}
\phi(u_0) &= P(M = 0) + \int_0^{u_0} P(H_1 \leq u_0| H_1=x) f_H(x) \ud x \\
&= (1-\theta) + \int_0^{u_0} P(\mathrm{age} \leq u_0 - x) f_H(x) \ud x.
\end{align*}
Since the ascending ladder process renews at $H_1$, the probability of the age of the residual process is $P(\mathrm{age} \leq u_0 - x) = \phi(u_0 - x)$. Thus $\phi(u_0) = P(M \leq u_0)$ satisfies the renewal equation $\phi(u_0) = (1-\theta) + \int_0^{u_0} \phi(u_0 - x) f_H(x) \ud x.$ 

\vspace{0.1cm}
(2) The event $\{M \leq u_0\}$ occurs if and only if $\max(Z_0, Z_0 + Z_1, Z_0+Z_1 + Z_2, \ldots) \leq u_0$. Conditioning on $Z_0 = y$, this is equivalent to 
\[ Z_0 = y \leq u_0 \quad \mathrm{and} \quad \max(0, Z_1, Z_1 + Z_2, \ldots) \leq u_0 - y. \]
Here, $P(\max(0, Z_1, Z_1 + Z_2, \ldots) \leq u_0 - y) = \phi(u_0 - y)$. De-conditioning over all possible $y$, we get $\phi(u_0) = \int_{-\infty}^{u_0} \phi(u_0 - y) f_Z(y) \ud y$, which by change of variable $x = u_0 - y$ becomes $\phi(u_0) = \int_0^\infty \phi(x) f_Z(u_0 - x) \ud x$.
\end{IEEEproof}

Equations (\ref{eqn:renewal-self-sus-1}) and (\ref{eqn:renewal-self-sus-2}) relate $\phi$ with $H$ (a defective random variable) and $Z$ (a proper random variable), respectively. Equation (\ref{eqn:renewal-self-sus-1}) can be recognized as a \tit{renewal equation} while (\ref{eqn:renewal-self-sus-2}) can be recognized as a Wiener-Hopf integral. The form of (\ref{eqn:renewal-self-sus-2}) also suggests the possibility of using an iterative solution procedure to obtain $\phi$.

\begin{corollary}
Given the self-sustainability condition, $\phi(u_0)$ is  a proper distribution with (i) $\phi(u_0) = 0$ for $u_0 < 0$,  (ii) $\phi(0) = 1 - \theta$, and (iii) $\phi(\infty) = 1$. 
\label{coro:phi-is-proper}
\end{corollary}
\begin{IEEEproof}
(i) Follows from the fact that $u_0$ only takes non-negative values. (ii) By putting $u_0 = 0$ in (\ref{eqn:renewal-self-sus-1}). (iii) Putting $u_0 = \infty$ in (\ref{eqn:renewal-self-sus-1}), we have $\phi(\infty) = (1-\theta) + \phi(\infty) \int_0^\infty f_H(x)\ud x$. Recalling that $H$ is defective, with $F_H(\infty) = \theta$, we have $\phi(\infty) = 1$. 
\end{IEEEproof}

\tbf{Remark:} \tbf{Corollary \ref{coro:phi-is-proper}} can also be proved as a consequence of \textbf{Proposition \ref{prop:selfsus-renewal-fun}}, since $\zeta(0) = 1$ and $\zeta(\infty) = 1/(1-\theta)$ when $H$ is defective. In the above corollary, it is remarkable that the system can be self-sustaining even when there is no initial battery energy.  In other words, this is the probability that the random walk $U_n$ starting from the origin will always be positive. This also allows us to interpret $\theta$ as the eventual outage probability when there is no initial battery energy, i.e. $\psi(0) = \theta$. The corollary also guarantees that as $u_0$ becomes large, the eventual outage becomes zero. Thus, a possible strategy in reducing the eventual outage is to simply increase the initial battery energy. We will later show that the rate at which the eventual outage decreases with $u_0$ is exponential. 

\tbf{Remark:} Equations (\ref{eqn:selfsus-renewal-fun}) and (\ref{eqn:renewal-self-sus-1}) can also be expressed in terms of Laplace transform. Taking the Laplace transform of (\ref{eqn:renewal-self-sus-1}), we have 
\begin{align}
\widehat{\phi}(r) &= \frac{1 - \theta}{r} +  \widehat{\phi}(r) \widehat{f}_H(r), \nonumber \\
\therefore \; \widehat{\phi}(r) &= \frac{1 - \theta}{r (1 - \widehat{f}_H(r))}.   \label{eqn:LT-of-self-sus}
\end{align}

The exact relationship between $H$ and $Z$ is given by the well known Weiner-Hopf factorization identity \cite[Ch XII.3]{Feller1971} \cite[Ch VIII.3]{Asmussen2003} in terms of their moment generating functions (MGFs) as:
\beq
1 - \Mcal_Z = (1 - \Mcal_H)(1 - \Mcal_{H_{-}}),
\label{eqn:weiner-hopf-mgf}
\eeq
where $H_{-}$ is the descending ladder height, defined in a manner similar to the ascending ladder height process. The $H_-$ is defined over $(-\infty,0]$. The above identity is also written in terms of convolution as
\beq
F_Z = F_H + F_{H_{-}} - F_{H}*F_{H_{-}}.
\label{eqn:weiner-hopf-conv}
\eeq

Likewise, the distributions of $Z$, $H$, and $H_{-}$ are related to the renewal function $\zeta$ by
\begin{subequations}
\begin{align}
F_{H}(x) &= \zeta_{-}*F_Z(x), & x > 0, \label{eqn:H-zeta-Za} \\
F_{H_{-}}(x) &= \zeta*F_Z(x), & x \leq 0 \label{eqn:H-zeta-Zb},
\end{align}
\end{subequations}
where $\zeta_{-}$ is the renewal function defined by $F_{H_{-}}$ in a manner similar to $\zeta$ given in (\ref{eqn:renewal-function}). 

The important thing to note here is that when $\Ebb[Z] < 0$, while $F_H$ is defective, $F_{H_{-}}$ is proper. Since $H_-$ is defined over $(-\infty,0]$, this gives us the condition that $F_{H_{-}}(0) = 1$. Lastly, since $H_-$ is a proper distribution, the descending ladder process is a proper renewal process.

The factorization (\ref{eqn:weiner-hopf-mgf}) is in itself difficult to perform explicitly. As such, we will focus on obtaining bounds and asymptotic approximations of $\phi$ (or equivalently, $\psi$).


\section{Bound on Eventual Outage Probability}
Thus far we have described the energy surplus process  and related the various associated concepts to the eventual outage$\slash$self-sustainability probability. These probabilities can be evaluated by solving the formulas given in\tbf{ Propositions~ \ref{prop:selfsus-renewal-fun}}~and~\tbf{\ref{prop:outage-convolution-formula}}. However, doing so is not trivial. As such, we wish for a simple bound to estimate the eventual outage probability.  Here we will establish a tight exponential bound for the eventual outage probability using the concept of martingales. 

\begin{definition}[Adjustment Coefficient]
The value $r^*\neq0$ is said to be the adjustment coefficient of $X$ if $\Ebb[\exp(r^* X)] = \int e^{r^* x} \ud F_X = 1$.
\label{def:adj-coefficient}
\end{definition}

\begin{definition}[Martingale]
A process $\{X_i\}$ is said to be a martingale if $\Ebb[X_{n+1}|X_n, \ldots, X_0] = X_n$.
\label{def:martingale}
\end{definition}

\begin{lemma}
Let $\{Z_i\}$, $\{S_i\}$ be as before. Suppose there exists an adjustment coefficient $r^* > 0$ such that $\Ebb[\exp(r^*Z_i)] = 1$, then $\exp(r^* S_n)$ for $n=0, 1, 2, \ldots$ is a martingale.
\end{lemma}
\begin{IEEEproof}
We have
\begin{align*}
& \Ebb[\exp(r S_{n+1})|S_n,\ldots,S_1] \\
= & \; \Ebb[\exp(r (S_n + Z_{n+1})) | S_n,\ldots,S_1] \\
= & \; \Ebb[\exp(r Z_{n+1})] \cdot \Ebb[\exp(r S_n)| S_n, \ldots, S_1] \\
= & \; \Ebb[\exp(r Z_{n+1})] \cdot \exp(r S_n).
\end{align*}
Since there exists a constant $r^* > 0$ such that $\Ebb[\exp(r^* Z_{n+1})] = 1$, then $\Ebb[\exp(r^* S_{n+1})|S_n,\ldots,S_1] = \exp(r^* S_n)$, satisfying the definition of a martingale.
\end{IEEEproof}

\tbf{Remark:} Note that since $Z_i$ is a proper random variable, it is trivially true that $\Ebb[\exp(r^*Z_i)] = 1$ if $r^* = 0$. In the following lemma, we will show that there exists a non-trivial value of $r^* > 0$ for which this property holds true. It will also give the conditions under which the adjustment coefficient will exist.

\begin{lemma}
Suppose that $\Ebb[Z_i] < 0$, which is the self-sustainability condition. Also, assume that there is $r_1 > 0$ such that the moment generating function (MGF) $\Mcal_Z(r) = \Ebb[e^{r Z_i}] < \infty$ for all $-r_1 < r < r_1$, and that $\lim_{r \rightarrow r_1} \Mcal_Z(r) = \infty$. Then, there is a unique adjustment coefficient $r^* > 0$.
\end{lemma}
\begin{IEEEproof}
Since the MGF exists in the neighborhood of zero, derivative of every order exists in $(-r_1, r_1)$. By definition, $\Mcal_Z(0) = 1$. Since $\Ebb[Z_i]<0$ by assumption, we have $\Mcal'_Z(0) = \Ebb[Z_i] < 0$, which means that $\Mcal_Z(r)$ is decreasing in the neighborhood of 0. Also, since $\Mcal''_Z(r) = \Ebb[Z_i^2 e^{rZ_i}] > 0$, due to the fact that the expectation of a positive random variable is positive, it follows that $\Mcal_Z$ is convex on $(-r_1, r_1)$. Again, by assumption, we have $\lim_{r \rightarrow r_1} \Mcal_Z(r) = \infty$. It now follows that there exists a unique $s \in (0, r_1)$ such that $\Mcal_Z(s) < 1$ and $\Mcal_Z'(s) = 0$; and that on the interval $(s,r_1)$ the function $\Mcal_Z(\cdot)$ is strictly increasing to $+\infty$. As such, there exists a unique $r^* \in (0, r_1)$ such that $\Mcal_Z(r^*) = 1$. Since the MGF does not exist on $[r_1, \infty)$, it follows that $r^*$ is unique on $(0,\infty)$.
\end{IEEEproof}

\tbf{Remark:} The above lemma characterizes the class of distributions of $Z$ for which the adjustment coefficient will exist. If any one of the required conditions is violated, then $r^*$ will not exist; and the subsequent results on the exponential bound and asymptotic approximation, which depends on $r^*$, will not be valid. For instance, if the size of the energy packet has a heavy tail distribution for which MGF does not exist (e.g. log-normal, Weibull, Pareto), then the MGF of $Z$ will also not exist; hence, $r^*$ will not exist for this case. For such cases, we may resort to solving the Weiner-Hopf integrals given in \tbf{Proposition~\ref{prop:outage-convolution-formula}}. 

In the following proposition, we will use the double barrier argument to bound the eventual outage probability.

\begin{proposition}
Assume that the self-sustainability condition $\Ebb[Z_i]  < 0$ holds, and that the adjustment coefficient $r^* > 0$ exists. Then the eventual outage probability is bounded by
\beq
\psi(u_0) \leq \exp(-r^* u_0), \quad u_0 > 0.
\label{eqn:outage-inequality}
\eeq
\label{prop:outage-inequality}
\end{proposition}
\begin{IEEEproof}
Put $\tau(u_0) = \inf(n \geq 0 : S_n > u_0)$. The $\tau(u_0)$ is the first passage time that the random walk $\{S_n\}$ exceeds $u_0 > 0$ in the positive direction. The event that $\{ \tau(u_0) \leq k \}$ is equivalent to the union of events $\cup_{i=0}^k \{S_i > u_0\}$. Similarly, for $a > 0$, let $\sigma(a) = \inf\{n \geq 0 : S_n < - a\}$.  Here $\sigma(a)$ denotes the first passage time that the random walk $\{S_n\}$ exceeds $ - a < 0$ in the negative direction. Since $\Ebb[Z_i]  < 0$, by Case~3 of Theorem~\ref{theorem:random_walk}, $P(\sigma(a) < \infty) = 1$. Hence, $(\tau(u_0) \wedge \sigma(a)) \equiv \inf(\tau(u_0),\sigma(a))$ is a stopping time with $P((\tau(u_0) \wedge \sigma(a))<\infty) = 1$. Since $\exp(r^* S_n)$ for $n=0,1,2, \ldots$ is a martingale, using the optional sampling theorem, we have
 \begin{align*}
 1 &= \Ebb[e^{r^*S_0}]  = \Ebb[\exp(r^* S_{(\tau(u_0)\wedge\sigma(a))})] \\
 &= \Ebb[\exp(r^* S_{\tau(u_0)}) | \tau(u_0) < \sigma(a)] \\&  \qquad \quad + \Ebb[\exp(r^* S_{\sigma(a)}) | \sigma(a) < \tau(u_0) ] \\
 &\geq \Ebb[\exp(r^* S_{\tau(u_0)}) | \tau(u_0) < \sigma(a)] \\
 &\geq e^{r^* u_0} P(\tau(u_0) < \sigma(a)) 
 \end{align*}
as $S_{\tau(u_0)} > u_0$ in the last step. Since $P(\lim_{a \rightarrow \infty} \sigma(a) = \infty) =1$, thus letting $a \rightarrow \infty$ we get $\psi(u_0) = P(\tau(u_0) < \infty) = \lim_{a \rightarrow \infty} P(\tau(u_0) < \sigma(a)) \leq e^{-r^* u_0}$, as required. 
\end{IEEEproof}

\tbf{Remark:} This proposition quantifies how fast the eventual outage probability diminishes with increasing initial battery energy. Thus, during the design of a system, where we are willing to tolerate an arbitrarily small eventual outage probability, our task is to determine the initial battery energy. We can use the above bound to roughly calculate the required initial battery energy. The corollaries below   follow immediately: 

\begin{corollary}
Assuming that the self-sustainability condition holds and $r^*>0$ exists, then  $\lim_{u_0 \rightarrow \infty} \psi(u_0) = 0.$
\end{corollary}

\begin{corollary}
Assuming that the self-sustainability condition holds and the adjustment coefficient $r^*>0$ exists, for a given tolerance $\epsilon \in (0,1)$, if the eventual outage probability is constrained at $\psi(u_0) = \epsilon$, then the maximum initial battery energy required is $u_0 = \frac{1}{r^*} \log (\frac{1}{\epsilon})$. 
\label{coro:initial-energy}
\end{corollary}

\begin{corollary}
Let the system restart with the same initial battery energy $u_0$ after an outage. If we constrain at most $k$ outages with probability $P(\mathrm{at\;most}\;k \; \mathrm{outages}) = 1-\epsilon$, where $\epsilon \in (0,1)$, then assuming the self-sustainability condition holds and the adjustment coefficient $r^* > 0$ exists, the initial battery energy required is at most $u_0 = \frac{1}{(k+1) r^*} \log(\frac{1}{\epsilon})$.
\label{coro:initial-energy-with-restarts}
\end{corollary}

\begin{IEEEproof}
From \tbf{Proposition~\ref{prop:no-of-outages}}, we have $P(\mathrm{at\;most}\;k \; \mathrm{outages}) = \sum_{i=0}^k \psi^i \phi = 1 - \psi^{k+1}$. Using this in the constraint $P(\mathrm{at\;most}\;k \; \mathrm{outages}) = 1-\epsilon$, we have $\log \psi = \frac{1}{k+1} \log(\epsilon)$. From the inequality $\log \psi \leq - r^* u_0$, we have the maximum required initial battery energy as $u_0 = \frac{1}{(k+1) r^*} \log (\frac{1}{\epsilon})$.
\end{IEEEproof}


\section{Asymptotic Approximation of Eventual Outage Probability}
While the exponential bound given in the previous section is simple to use, we can sharpen our estimates using the key renewal theorem for defective distribution. The basic idea behind this approach is to transform a defective distribution into a proper distribution using the adjustment coefficient, and then apply the key renewal theorem for proper distribution. 

\subsection{Asymptotic Approximation}
First, we will define the renewal equation and then give the related theorem.

\begin{definition}
The renewal equation is the convolution equation for the form $Z = z + F*Z$, where $Z$ is an unknown function on $[0,\infty)$, $z$ is a known function on $[0,\infty)$ and $F$ is a known non-negative measure on $[0,\infty)$. Often $F$ is assumed to be a probability distribution. If $F(\infty) = 1$, then the renewal equation is proper. If $F(\infty) < 1$, then the renewal equation is defective. 
\end{definition}

\begin{theorem}\cite[Ch. V, Prop 7.6, p. 164]{Asmussen2003} \cite[Ch. IX.6, Theo. 2, p. 376]{Feller1971}
Suppose that for defective distribution $F$, there exists an adjustment coefficient $r^* > 0$ such that $\tilde{\mu} = \int t e^{r^* t} dF < \infty$ exists. If in the defective renewal equation $Z = z + F*Z$, $z(\infty) = \lim_{t\rightarrow\infty}z(t)$ exists and $e^{r^* t}(z(t) - z(\infty))$ is directly Riemann integrable, then the solution of the renewal equation satisfies
\beq
\tilde{\mu} e^{r^* t} [Z(\infty) - Z(t)] \sim \frac{z(\infty)}{r^*} + \int_0^\infty e^{r^*s} [z(\infty)-z(s)]\ud s. 
\label{eqn:key-renewal-eqn-defective}
\eeq
\label{theorem:key-renewal-eqn-defective}
\end{theorem}

\begin{proposition}
Given the self-sustainability condition, the eventual outage probability is asymptotically given by
\beq
\psi(u_0) \sim \frac{1-\theta}{r^* \tilde{\mu}_H}  e^{-r^* u_0},
\label{eqn:asymp-eventual-outage}
\eeq
where $r^*$ is the adjustment coefficient of $H$ and $\tilde{\mu}_H = \int x e^{r^* x} f_H(x) \ud x < \infty$.
\label{prop:asymp-eventual-outage}
\end{proposition}
\begin{IEEEproof}
Recall that (\ref{eqn:renewal-self-sus-1}) is in the form of a renewal equation with the  defective distribution of $H$. As such, in \textbf{Theorem~\ref{theorem:key-renewal-eqn-defective}}, we have $z(t) \equiv 1 - \theta$ and $Z \equiv \phi(u_0)$. Thus, the integral in (\ref{eqn:key-renewal-eqn-defective}) vanishes and
\[ \tilde{\mu}_H e^{r^* u_0} [\phi(\infty) - \phi(u_0)] \sim \frac{1 - \theta}{r^*}. \]
From \textbf{Corollary \ref{coro:phi-is-proper}}, we know that $\phi$ is proper. Thus $\phi(\infty) - \phi(u_0) = 1 - \phi(u_0) = \psi(u_0)$. Hence, we have the desired result.
\end{IEEEproof}

Note that the adjustment coefficient $r^*$ for $H$ is the same as that for $Z$. This is easily demonstrated, since $\Mcal_Z(r^*) = \Ebb[e^{r^*z}] =  1$ where $r^* > 0$, we have from the Weiner-Hopf factorization (\ref{eqn:weiner-hopf-mgf}), 
\[ (1 - \Mcal_H(r^*))(1 - \Mcal_{H_{-}}(r^*)) = 1 - \Mcal_Z(r^*) =  0. \]
Here, since $H_-$ is defined over $(-\infty,0]$, we have 
\begin{align*}
\Mcal_{H_-}(r^*) &= \int_0^\infty e^{-r^* x} f_{H_-}(-x) \ud x, \\
 & \stackrel{(a)}{<} \int_0^\infty f_{H_-}(-x) \ud x,  \\
 & \stackrel{(b)}{=} 1, 
\end{align*} 
since in $(a)$ $r^* > 0$ and $f_{H_-}$ is positive, while in $(b)$ $H_{-}$ is a proper distribution. Thus, $1 - \Mcal_{H_{-}}(r^*) >0$, and this implies that  $\Mcal_H(r^*) = 1$, making $r^*$ the adjustment factor of $H$ as well.

The above result is asymptotic in the sense that higher values of $u_0$ give more accurate results. However, without further modeling assumptions, this is as far as we can proceed, since an explicit evaluation of $F_H$ is difficult. In \tbf{Section~\ref{sec:specialcase}}, we will explore the case when the energy packet arrival is modeled as a Poisson process.


\subsection{Computing the Adjustment Coefficient}
The computation of the adjustment coefficient $r^*$ is not trivial. As given by \tbf{Definition~\ref{def:adj-coefficient}}, the adjustment coefficient must satisfy the condition $\Ebb[e^{r^*Z}] = \Mcal_Z(r^*) = 1$. For the computational purpose, it is more convenient to use the cumulant generating function (CGF) of $Z$ rather than its MGF. The CGF of $Z$ is defined as $\Kcal_Z(r) = \log \Mcal_Z(r)$. In terms of CGF, the adjustment coefficient satisfies the condition $\Kcal_Z(r^*) = \log \Mcal_Z(r^*) = \log 1 = 0$. Hence, we see that the adjustment coefficient $r^*$ is a real positive root of CGF $\Kcal_Z(r)$, and can be found by solving the equation
\beq 
\Kcal_Z(r^*) = 0. 
\label{eqn:adj-coef-eqn}
\eeq
Since $Z_i = p A_i - X_i$ with $X_i$ and $A_i$ independent of each other, we can invoke the linearity of CGFs to obtain, $\Kcal_Z(r) = \Kcal_A(pr) + \Kcal_X(-r)$. Thus, (\ref{eqn:adj-coef-eqn}) becomes
\beq
 \Kcal_A(pr^*) + \Kcal_X(-r^*) = 0.
 \label{eqn:adj-coef-eqn-2}
\eeq

For the important case when the energy packet arrival is a Poisson process, the inter-arrival time $A_i \sim \mathrm{Exp}(\lambda)$. As such, $\Mcal_A(r) = \lambda/(\lambda - r)$, and $\Kcal_A(r) = - \log(1 - \frac{r}{\lambda})$. Thus, for Poisson arrivals, equation (\ref{eqn:adj-coef-eqn-2}) becomes $-  \log\left(1 - \frac{pr^*}{\lambda}\right) + \Kcal_X(-r^*) = 0$, which after exponentiation can be expressed as
\beq
1 - \frac{pr^*}{\lambda} = \Mcal_X(-r^*).
\label{eqn:adj-coef-eqn-MG1}
\eeq

Given the Poisson arrival, some possible distributions for the energy packets sizes and their corresponding solutions are as follows:
\begin{enumerate}
\item Assuming that $X_i$ are exponentially distributed, $X_i \sim \mathrm{Exp}(1/\bar{X})$, then we have $\Mcal_X(r) = 1/(1 - r\bar{X})$. Therefore, the solution to (\ref{eqn:adj-coef-eqn-MG1}) is 
\beq
r^* = \frac{1}{\bar{X}}\left(\frac{\lambda \bar{X}}{p} - 1\right).  
\label{eqn:adj-coef-eqn-MM1}
\eeq

\item Assuming $X_i$ are deterministic, $X_i = c$, then we have $\Mcal_X(r) = e^{cr}$. Therefore, the solution to (\ref{eqn:adj-coef-eqn-MG1}) is obtained by solving the equation
\beq
1 - \frac{pr}{\lambda} - e^{-cr} = 0.
\label{eqn:adj-coef-eqn-MD1}
\eeq

\item Assuming $X_i$ are uniformly distributed, $X_i \sim U(0,2\bar{X})$, then we have $\Mcal_X(r) = \frac{e^{2 \bar{X} r}-1}{2 \bar{X} r}$. Therefore, the solution to (\ref{eqn:adj-coef-eqn-MG1}) is obtained by solving the equation
\beq
e^{-2\bar{X} r} - \frac{2p\bar{X}}{\lambda} r^2 + 2 \bar{X} r + 1 = 0.
\label{eqn:adj-coef-eqn-MU1}
\eeq

 \item Assuming $X_i$ are chi-squared distributed, $X_i \sim \mathrm{\chi^2}(\bar{X})$, then we have $\Mcal_X(r) = (1 - 2r)^{-\bar{X}/2}$. So the solution to (\ref{eqn:adj-coef-eqn-MG1}) is obtained by solving the equation
\beq
1 - \frac{pr}{\lambda}  - \left(1 + 2r \right)^{-\bar{X}/2} = 0.
\label{eqn:adj-coef-eqn-MChi21}
\eeq

\item Assuming $X_i$ are inverse Gaussian distributed, $X_i \sim \mathrm{InvGauss}(\mathrm{\bar{X},\bar{X}^2})$, then we have $\Mcal_X(r) = \exp[\bar{X}(1 - \sqrt{1 - 2 r})]$. So the solution to (\ref{eqn:adj-coef-eqn-MG1}) is obtained by solving 
\beq
1 - \frac{pr}{\lambda}  - \exp[\bar{X}(1 - \sqrt{1 + 2 r})] = 0.
\label{eqn:adj-coef-eqn-MInvGauss1}
\eeq
\end{enumerate}

A simple approximation of $r^*$ can be obtained by making a formal power expansion of $\Mcal_X(-r)$ in terms of the moments of $X$ up to second order term as $\Mcal_X(-r) \approx 1- \bar{X} r + \frac{\overline{X^2}}{2}{r}^2$. Using this expression in (\ref{eqn:adj-coef-eqn-MG1}) and solving for $r^*>0$, we obtain
\beq
r^* \approx \frac{2p}{\lambda \overline{X^2}} \left( \frac{\lambda \bar{X}}{p} - 1 \right).
\eeq
The truncation error in the expansion of $\Mcal_X$ is $E = \sum_{i=3}^\infty \frac{\mu_i(X)}{i!} (-r)^i$, in which  $\mu_i(X)$ is the $i$-th moment of $X$. Since $X$ is a non-negative random variable, all of its moments will be positive. Let $K = \max_{i \geq 3} \mu_i(X)$ be the largest moment of order greater than 2. Then we have the bound on truncation error as $E \leq K \sum_{i=3}^\infty \frac{(-r)^i}{i!} < K e^{-r}$. From this inequality, we see that the effect of error is small when the actual value of $r^*$ is large. 

If instead, we have non-Poisson arrival, and supposing that the distribution of energy packet size is exponential, $\mathrm{Exp}(1/\bar{X})$, then (\ref{eqn:adj-coef-eqn-2}) becomes 
\beq
1 + \bar{X} r^* = \Mcal_A(pr^*).
\label{eqn:adj-coef-eqn-GM1}
\eeq
As before, expanding $\Mcal_A(pr)$ up to second order term and solving for $r^*>0$ gives the approximation
\beq
r^* \approx \frac{2}{\lambda p \overline{A^2}} \left(\frac{\lambda\bar{X}}{p} - 1\right).
\eeq

More generally, since $\Kcal_Z(r)$ can be expanded in terms of the mean and variance of $Z$ as $\Kcal_Z(r) \approx \mu_Z r + \frac{\sigma_Z^2}{2} r^2$, we have the approximate solution $r^*$ for (\ref{eqn:adj-coef-eqn}) as
\beq
r^* \approx - \frac{2\mu_Z}{\sigma_Z^2}.
\eeq
Since $\mu_Z = \Ebb[Z]  < 0$, the above approximation will correctly give $r^* > 0$. This value can be used as an initial point for a root finding algorithm. Better approximations can be found by including higher order terms in the expansion and reverting the series using Lagrange inversion.


\section{Special Case: Evaluation of Eventual Outage Probability for Poisson Arrivals}
\label{sec:specialcase}

In general, the density of $Z_i = pA_i - X_i$ is given in terms of the densities of $A_i$ and $X_i$ as
\beq
f_Z(z) = \frac{1}{p} \int_{\max(0,-z)}^\infty f_A\left(\frac{z+x}{p}\right) f_X(x) \ud x. 
\label{eqn:general-pdf-Z}
\eeq
When the energy packet arrival is assumed to be a Poisson process, the inter-arrival time $A_i$ is exponentially distributed, $A_i \sim \mathrm{Exp}(\lambda)$. As such, the density of $Z_i$ is 
\beq 
f_Z(z) = \frac{\lambda}{p} \; e^{-\frac{\lambda z}{p}}  \int_{\max(0,-z)}^\infty e^{-\frac{\lambda x}{p}} f_X(x) \ud x. 
\label{eqn:pdf-Z-when-A-exp}
\eeq

When $z \geq 0$, the density of $Z$ has the form
\[ f_Z(z) = \frac{\lambda}{p} \; e^{-\frac{\lambda z}{p}}  \int_0^\infty e^{-\frac{\lambda x}{p}} f_X(x) \ud x, \qquad z \geq 0.  \]
Since the above integral is independent of $z$, the right tail of the density is exponential. That is, $f_Z(z) = C e^{-\frac{\lambda z}{p}}$ for $z\geq0$, where $C$ is some constant given by $C = \frac{\lambda}{p} \int_0^{\infty} e^{-\frac{\lambda x}{p}} f_X(x) \ud x$. Now, from (\ref{eqn:H-zeta-Za}), we have \[ f_H(x) = \zeta_{-}*f_Z(x) = C \int_{-\infty}^0 e^{-\frac{\lambda (x - s)}{p}} \zeta_{-}(s) \ud s. \]
Again we see that regardless of the expression for $\zeta_{-}$, $f_H$ takes an exponential form given by
\[ f_H(x) = K e^{-\frac{\lambda x}{p}}, \]
where $K = C  \int_{-\infty}^0 e^{\frac{s}{p}} \zeta_{-}(s) \ud s$. 

Since we know that $H$ is defective, multiplying $f_H(x)$ by the adjustment factor $e^{r^*x}$ should convert it into a proper distribution. From the normalization condition for proper distributions, $\int_0^\infty e^{r^* x} f_H(x) \ud x = 1$, we can solve for $K = \frac{\lambda}{p} - r^*$. Thus, we can re-write $f_H$ as
\[ f_H(x) = \left(\frac{\lambda}{p} - r^* \right) e^{-\frac{\lambda x}{p}},\]
such that $\theta = F_H(\infty) = 1 - \frac{r^* p}{\lambda}$. Hence, the amount of defect is $1 - \theta = \frac{r^* p}{\lambda}$. When $f_H(x)$ is multiplied by $e^{r^* x}$, we obtain the proper distribution 
\[ e^{r^* x} f_H(x) = \left(\frac{\lambda}{p} - r^* \right) e^{-(\frac{\lambda}{p} - r^*) x}, \]
and the mean of this proper exponential distribution is $\tilde{\mu}_H = \left( \frac{\lambda}{p} - r^* \right)^{-1}$. Thus, we have from  (\ref{eqn:asymp-eventual-outage}) of \tbf{Proposition \ref{prop:asymp-eventual-outage}},
\beq 
\psi(u_0) \sim \left(1 - \frac{r^* p}{\lambda}\right) e^{-r^* u_0}. 
\label{eqn:asymp-MG1}
\eeq

When $X$ is also exponentially distributed, we have the exact value of $r^*$ from (\ref{eqn:adj-coef-eqn-MM1}). Hence, we have 
\beq 
\psi(u_0) \sim  \frac{p}{\lambda \bar{X}}e^{-r^* u_0}. 
\label{eqn:asymp-MM1}
\eeq
In fact, the (\ref{eqn:asymp-MG1}) and (\ref{eqn:asymp-MM1}) are not just asymptotic approximations, but also exact formulas (see \cite[Ch XII.5, Ex 5(b)]{Feller1971}). From these arguments, we have the following proposition:

\begin{proposition}
Assume that the self-sustainability condition holds and the adjustment coefficient $r^*>0$ exists. If the  energy packets arrive  into an HSC system as a Poisson process, then the eventual outage probability is given by 
\beq
\psi(u_0) = \left(1 - \frac{r^* p}{\lambda}\right) e^{-r^* u_0}. 
\label{eqn:eventual-outage-poisson-exact}
\eeq
Furthermore, if the energy packet size is also exponentially distributed, then 
\beq
\psi(u_0) = \frac{p}{\lambda \bar{X}} \exp\left\{{-\frac{1}{\bar{X}}\left(\frac{\lambda \bar{X}}{p} - 1\right) u_0}\right\}. 
\label{eqn:eventual-outage-poisson-exponential-exact}
\eeq
\end{proposition}
\begin{IEEEproof}
From discussion above.
\end{IEEEproof}


\section{Battery Energy Evolution Process}
\label{sec:battery-energy-evolution}
So far we have directed our attention to the energy surplus $U(t)$ and the case when the self-sustainability condition is satisfied. For the sake of completeness, let us now consider the battery energy $W(t)$ at time $t$ and the case when the self-sustainability condition is not satisfied. 

\subsection{Equivalence with Queueing Systems}
We will first prove that the battery energy process is a Lindley process. Making this identification will then allow us to compare the HSC system to a $GI / G / 1$ queue, which in turn will allow us to exploit the results from queueing theory, for which the Lindley process was first studied. When the self-sustainability condition is not satisfied, the battery energy process $W(t)$ is stationary and ergodic. Thus, it makes better sense to talk about the outage probability $P(W(t) = 0)$ of the system rather than the eventual outage probability which is always unity, i.e. $\psi(u_0) = 1$.

\begin{definition}\cite[Ch 3.6]{Asmussen2003}
A discrete-time stochastic process $\{Y_i\}$ is a Lindley process if and only if $\{Y_i\}$ satisfies the recurrence relation 
\beq 
Y_{n+1} = \max(0, Y_n + X_n), \quad n = 0,1, \ldots 
\label{eqn:lindley-recursion}
\eeq
where $ Y_0 = y \geq 0$ and $\{X_i\}$ are independent and identically distributed. This recursive equation is called Lindley recursion.\footnote{The Lindley process is referred to as queueing process in \cite{Feller1971}.}
\end{definition}

\begin{lemma}
The battery energy process $W(t)$ observed just before the of arrival energy packets  is a Lindley process and satisfies the recursion $W_{n+1} = \max(0,W_n + Z'_n)$, for $n=0,1,\ldots,$ where $W_0 = u_0$ and $Z'_n = - Z_n$.
\label{lemma:HST-Lindley}
\end{lemma}
\begin{IEEEproof}
As with the energy surplus in the previous section, let $W_n$ be the amount of battery energy immediately before the arrival of $n$-th energy packet. The initial battery energy immediately before the arrival of the first energy packet $E_0$ is $W_0 = u_0$. The amount of battery energy just before the arrival of next $(n+1)$-th energy packet, $W_{n+1}$, is then the sum of $W_n$ and $X_n$, the amount of energy contributed by the $n$-th packet into the battery, minus the amount consumed during the inter-arrival period of the $(n+1)$-th packet, $p A_n$. Thus we have $W_{n+1} = W_n + X_n - p A_n$ if $W_n + X_n - p A_n \geq 0.$ Likewise, the battery will be empty, $W_{n+1} = 0$, if $W_n + X_n - p A_n \leq 0$. Putting both of them together, we have 
 \beq
 W_{n+1} =\left\{ \begin{array}{lcr}
 			W_n + Z'_n, & \mathrm{if} & W_n + Z'_n \geq 0 \\
 			0, & \mathrm{if} & W_n + Z'_n \leq 0 .
 			\end{array} \right.
 \label{eqn:batt-energy}
 \eeq
 where $Z'_n = X_n - p A_n$. Since $\{X_n\}$ and $\{A_n\}$ are independent and identically distributed, $\{Z'_n\}$ is independent and identically distributed as well. We can write (\ref{eqn:batt-energy}) in compact form as  $W_{n+1} = \max(0,W_n + Z'_n)$, which is a Lindley recursion as given in (\ref{eqn:lindley-recursion}). Since $\{W_i\}$ satisfies the Lindley recursion, $\{W_i\}$ is a Lindley process. 
\end{IEEEproof}

The fact that $\{W_i\}$ is a Lindley process gives rise to a number of important consequences, as stated in the following propositions.

\begin{proposition}
The HSC system is equivalent to a $GI / G / 1$ queueing system.
\label{prop:HSC-GG1-equivalence}
\end{proposition}
\begin{IEEEproof}
The proposition follows from the fact that the virtual waiting time (i.e. the amount of time the server will have to work until the system is empty, provided that no new customers arrive, or equivalently, the waiting time of a customer in a first-in-first-out queueing discipline) of an $n$-th customer arriving into a $GI/G/1$ queueing system is a Lindley process \cite[Ch 3.6, Ex 6.1]{Asmussen2003}. Since HSC system is also a Lindley process by Lemma \ref{lemma:HST-Lindley}, the equivalence is established.
\end{IEEEproof}

\begin{proposition}
The  HSC system is ergodic and stationary if and only if $\Ebb[Z_i] > 0$, i.e. $\lambda \bar{X} < p$, when the self-sustainability condition is not satisfied. Under this condition, there will exist a unique stationary distribution for $W_n$, independent of the initial condition $W_0$, which is given by the Lindley's integral equation:
 \beq 
 F_W (w) = \int_{0-}^\infty F_Z(w-x) \ud F_W(x), \qquad w \geq 0  
 \label{eqn:steady-state-W}
 \eeq
\end{proposition}

\begin{IEEEproof}
This is a standard result from queueing theory for $GI/G/1$ queues, which by \tbf{Proposition \ref{prop:HSC-GG1-equivalence}} also applies to HSC system. See \cite[Coro 6.6]{Asmussen2003}.  
\end{IEEEproof}

\tbf{Remark:} We see that when the self-sustainability condition is not satisfied, the battery energy process of the system is ergodic and stationary. The equation (\ref{eqn:steady-state-W}) is again a Weiner-Hopf integral; and except for some special cases, its general solution is difficult to obtain. Nevertheless, it allows us to compute the outage probability, $P_{out} = P(W(t) = 0) = F_W(0)$, by invoking the ergodicity and stationarity of $W(t)$. 

\tbf{Remark:} The previous proposition also shows the logical connection between the eventual outage probability, $\psi$, and the outage probability, $P_{out}$. The eventual outage probability is 1 if and only if the outage probability is non-zero (i.e. $\psi = 1 \Leftrightarrow P_{out} > 0$). Likewise, the eventual outage probability is less than 1 if and only if the outage probability is zero (i.e. $\psi < 1 \Leftrightarrow P_{out} = 0$). 

\begin{table}
  \caption{Queueing Analogue}
  \centering
 \begin{tabular}{cll}
 \tbf{Parameter} & \tbf{Energy Harvesting} & \tbf{Queueing} \\
 \hline \\
  - & Consumer & Server \\
  - & Battery & Buffer \\
 $E_n$ & $n$-th energy packet & $n$-th customer \\
  $\lambda$ & Packet arrival rate & Customer arrival rate \\
 $W_n$ & Battery energy  & Virtual waiting time / work load \\ 
 $X_n$ & Packet size & Service time  \\ 
 $p A_n$ & Energy consumed & Inter-arrival time (scaled)  \\
 $P(W(t) = 0)$ & System outage & System idle \\
 -  & Self-sustainability & System always busy \\ 
 $\rho = \frac{\lambda \bar{X}}{p}$ & Utilization factor & Utilization factor (traffic intensity)
 \end{tabular}
 \label{table:queueing_analogue}
 \end{table}

While it generally difficult to find the distribution of $W$ from (\ref{eqn:steady-state-W}), nevertheless we have can find the probability that the battery is empty, $W=0$.

\begin{proposition}
\label{prop:outage-prob-formula}
For the HSC system, assuming that the self-sustainability condition is not satisfied, i.e. $\lambda \bar{X} < p$, let $\rho = \frac{\lambda \bar{X}}{p}$. Then, the outage probability is 
\beq
P_{out} = 1 - \rho.
\label{eqn:outage-prob-formula}
\eeq
\end{proposition}
\begin{IEEEproof}
From \tbf{Proposition \ref{prop:HSC-GG1-equivalence}}, the HSC system is equivalent to a $GI/G/1$ queuing system, with the battery energy corresponding to the virtual waiting time and $\rho$ corresponding to the utilization factor (see \tbf{Discussion 1} and \tbf{Table \ref{table:queueing_analogue}} for more). Accordingly, from \cite[Ch. X, Eqn 3.2]{Asmussen2003},  for a $GI/G/1$ queue when $\rho < 1$, the probability of virtual waiting time being zero is $1 - \rho$. Thus, from our correspondence we have $P_{out} = P(W = 0) = 1 - \rho$.
\end{IEEEproof}

\subsection{Discussion}
\begin{enumerate}
\item By comparing the Lindley recursion equations for an energy harvesting system and a queueing system, we can translate the terms and concepts of one system into those of the other. The Lindley recursion for a $GI/G/1$ queueing system is given by $V_{n+1} = \max(0, V_n + X_n - T_n)$, where for an $n$-th customer $V_n$ is its virtual waiting time, $X_n$ is its service time, and $T_n$ is the inter-arrival time between customers $n$ and $n+1$. In Table \ref{table:queueing_analogue}, we compare the terminologies of our energy harvesting system with the terminologies of a queueing system. Equivalently, the units of energy in (\ref{eqn:batt-energy}) can be converted into units of time by dividing both sides of (\ref{eqn:batt-energy}) by $p$.

\item This comparison also means that we are justified in using the Kendall notation to refer to the different arrival processes, energy packet size distributions, and the number of consumers in an HSC system. An HSC system where the energy packets arrive as a Poisson process can be expressed in Kendall notation as an $M/G/1$ system. Furthermore, if the distribution of the energy packet size is exponential, then we have an $M/M/1$ system. If the energy inter-arrival is general distribution and the energy packet size is exponential, then we have an $GI/M/1$ system. Likewise, for deterministic arrivals, we have a $D/G/1$ system.

\item The consumer of our energy harvesting system is equivalent to the server of a traditional queueing system. Thus the figure of merit for an energy harvesting system, as given by the outage probability, corresponds to the probability of the server being idle. Likewise, the self-sustainability of the energy harvesting system is translated as the probability that the server always  remains busy.

\item However, one crucial difference is that, unlike the customers in a queue, the energy packets are ``anonymous'', in the sense that they lose their distinction once they enter into the battery. As such, it does not make any physical sense to talk about the number of packets in the system, in contrast to the number of customers in the system, except perhaps for mathematical convenience. Similarly, the queuing disciplines like first-in-first-out also lose their relevance in the energy harvesting framework.

\item When the self-sustainability condition is violated, it makes better sense to study the duration of outage and coverage rather than the eventual outage. Under steady state, the distribution of the outage duration is given by the residual time distribution of the underlying renewal process: $F_O(x) = \lambda \int_0^x [1 - F_A(s)] \ud s$. For Poisson arrival case, the coverage duration can also be obtained by solving Takacs' equation $\Mcal_C(r) = \Mcal_{\frac{X}{p}}(r + \lambda - \lambda \Mcal_C(r))$; thus the average outage duration is $\Ebb[O] = 1/\lambda$ while the average coverage duration is $\Ebb[C] = \frac{\bar{X}}{p(1-\rho)} = \frac{\rho}{\lambda(1-\rho)}$. Hence, the duty cycle of the system is $\frac{\Ebb[C]}{\Ebb[C] + \Ebb[O]} = \rho$.

\item When the self-sustainability condition is not satisfied, but $\lambda \bar{X} \approx p$, we can use the heavy traffic approximation. We have the tail probability of battery energy given by the Kingman's bound as $P(W \geq w) \leq e^{-r^* w}$, where $r^* > 0$ such that $\Mcal_Z(-r^*) = 1$. Also, the distribution of $W$ is given by $F_W(w) \approx 1 - \exp\left( - \frac{2 p (1-\rho)}{\lambda (p^2 \sigma_A^2 + \sigma_X^2)} w \right)$.
\end{enumerate}


\section{Numerical Verification}

\begin{figure}
\begin{center}
	\includegraphics[width=3.75in]{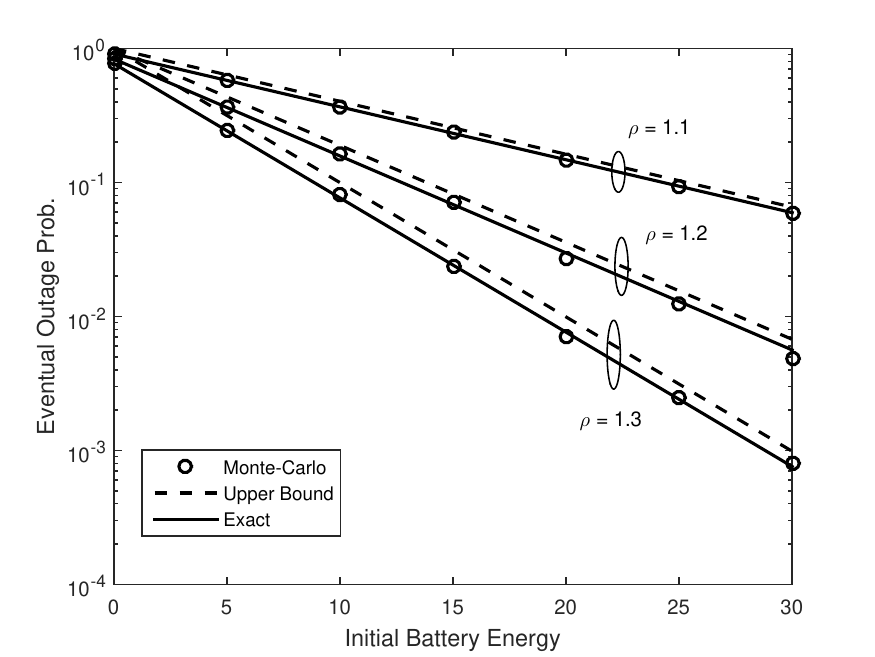}
	\caption{$\psi(u_0)$ versus $u_0$, when energy arrival is Poisson process and the energy packet size is exponentially distributed.}
	\label{fig:eventual-outage-for-exponential}
 \end{center}
\end{figure}

\begin{figure}
\begin{center}
	\includegraphics[width=3.75in]{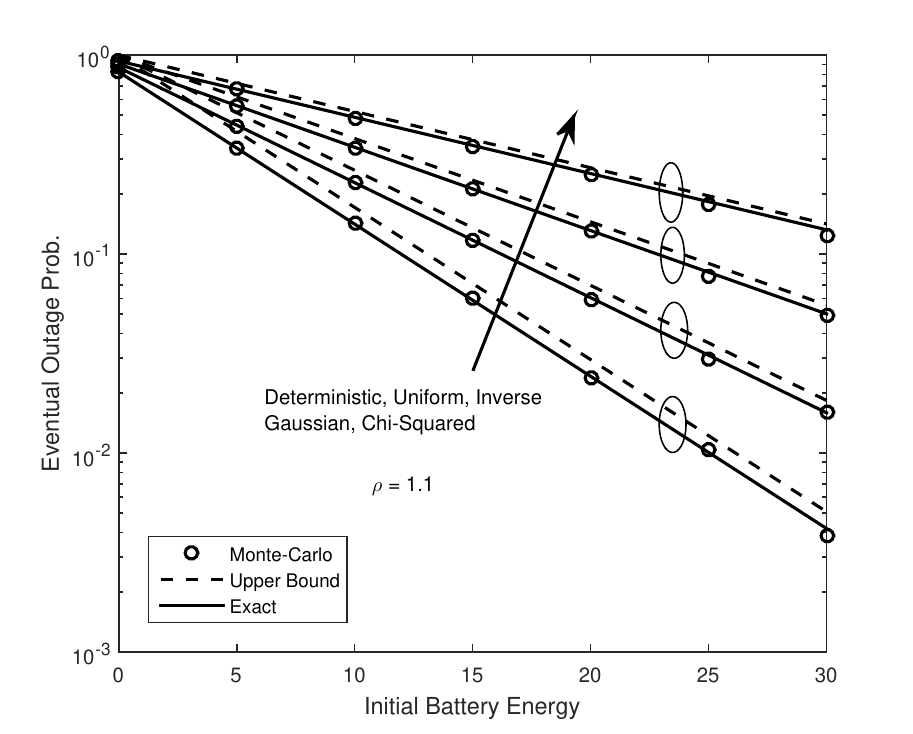}
	\caption{Comparison between eventual outage probabilities for various energy packet size distributions when $\rho = 1.1$.}
	\label{fig:eventual-outage-comparison-poisson-arrival}
 \end{center}
\end{figure}

\begin{figure}
\begin{center}
	\includegraphics[width=3.75in]{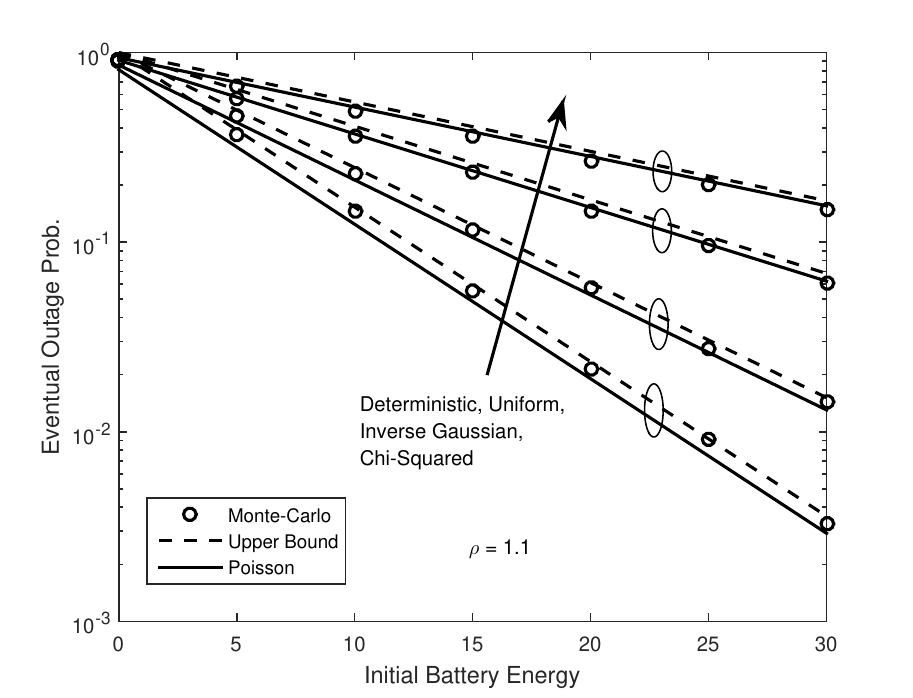}
	\caption{Comparison between eventual outage probabilities for various energy inter-arrival distributions when $\rho = 1.1$.}
	\label{fig:eventual-outage-comparison-non-poisson-arrival}
 \end{center}
\end{figure}

\begin{figure}
\begin{center}
	\includegraphics[width=3.75in]{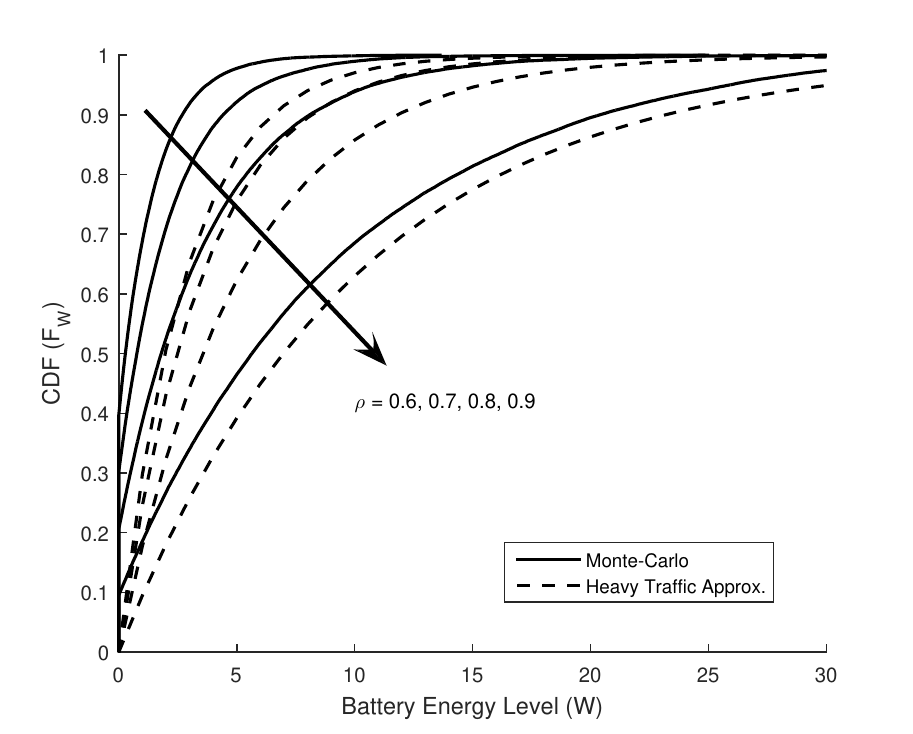}
	\caption{Distribution of battery energy when $\rho < 1$, for various values of $\rho$, compared with heavy traffic approximation.}
	\label{fig:battery-energy-cdf}
 \end{center}
\end{figure}

In this section, we verify the obtained formulas and bounds for the eventual outage probability. Here $\rho = \frac{\lambda \bar{X}}{p}$ is the utilization factor, in analogy with the queuing theory (see Table \ref{table:queueing_analogue}), which is a dimensionless number. For instance, $\rho=1.1$ would mean that the energy harvest rate is $10$ \% higher than the energy consumption rate. For Monte-Carlo simulations, we are required to follow the evolution of $U(t)$ until $t=\infty$, in order to check whether $U(t)$ becomes negative within finite time. However, this is certainly not possible. As such, we check the evolution of $U(t)$ through $t \in [0, T]$ where $T = 1000$ time-units. For a given value of $\rho$ and $u_0$ in some energy-units, $50,000$ sample paths were run to produce a single data point. 

The rationale behind taking a fixed large cut off value of time, $T$, as a substitute for infinity is that, even though we may not know the form of the distribution of the first outage time, $\tau < \infty$, the probability that an outage will occur beyond the cut off time $T$ is given by the tail probability of the first outage time distribution, $P(\tau > T| \tau < \infty)$. This tail probability becomes smaller as the cut off value $T$ becomes larger, i.e. $\lim_{T\to\infty} P(\tau > T| \tau < \infty) = 0$. What this means for the simulation is that if an outage has not been encountered within the given cut off value, then it is very unlikely that an outage will occur beyond this cut off. That is, since we have the unconditional statement $P(\tau > T) = P(\tau > T | \tau < \infty) P(\tau < \infty) + P(\tau = \infty)$, where for large cut off value $P(\tau > T | \tau < \infty) \approx 0$, we have $P(\tau > T) \approx P(\tau = \infty)$. Here, $P(\tau = \infty)$ is the probability that an outage will never occur. Since by definition the self-sustainability probability is $\phi = P(\tau = \infty)$, we have the approximation $\phi \approx P(\tau > T)$ for large $T$. Hence, in our simulation we are justified in counting the frequency of sample paths that do not undergo an outage until $T$ and taking it as the self-sustainability probability. Likewise, $\psi \approx 1 - P(\tau > T)$.

Figs. \ref{fig:eventual-outage-for-exponential} plots the eventual outage probability, $\psi(u_0)$, versus initial battery energy, $u_0$, for cases when the energy arrival is given by Poisson process and the energy packet size is exponentially distributed. The eventual outage probability is predicted using (\ref{eqn:eventual-outage-poisson-exponential-exact}). The adjustment coefficient $r^*$ is computed using (\ref{eqn:adj-coef-eqn-MM1}). The exponential upper bound is computed by using (\ref{eqn:outage-inequality}). In the semi-log plot, the eventual outage probability decreases linearly with respect to the initial battery energy. Also, the slope of the lines tend to become steeper with increasing $\rho$. This means that we require smaller initial battery energy for a given eventual outage probability, when the value of $\rho$ is higher. The results from Monte-Carlo simulations agree closely with that predicted by (\ref{eqn:eventual-outage-poisson-exact}). The upper bound given by (\ref{eqn:outage-inequality}) is observed to be tighter for smaller values of $\rho$. We also observe that the line for upper bound tends to run parallel to the exact line.

Fig. \ref{fig:eventual-outage-comparison-poisson-arrival} compares how the eventual outage probability changes with initial battery energy when the energy packet sizes is governed by deterministic, uniform, chi-squared, and inverse Gaussian distributions. In this figure, $\rho= 1.1$ and the energy arrival is a Poisson process for all the cases. The $r^*$ was obtained by numerically solving  (\ref{eqn:adj-coef-eqn-MD1}) -- (\ref{eqn:adj-coef-eqn-MInvGauss1}) for the respective cases, and the  eventual outage probability was predicted using (\ref{eqn:eventual-outage-poisson-exact}). We see that the trends are similar to the case with exponentially distributed energy packet size. We also see that the deterministic packet size gives the best performance, while the chi-squared distributed packet size performs the worst. This means that to achieve the same grade-of-service, say $\psi(u_0) = 0.01$, higher initial battery energy is required for the chi-squared case than for the deterministic case. 

Fig. \ref{fig:eventual-outage-comparison-non-poisson-arrival} compares how the eventual outage probability changes with initial battery energy when the energy inter-arrival times is governed by deterministic, uniform, chi-squared, and inverse Gaussian distributions, making the energy arrival process a non-Poissonian renewal process. In this figure, $\rho= 1.1$ and the energy packet size is assumed to be exponentially distributed for all the cases. The $r^*$ was obtained by numerically solving  (\ref{eqn:adj-coef-eqn-GM1}) for the respective cases. The MGFs for inter-arrival time, $A_i$, is similar to that given for $X_i$ in (\ref{eqn:adj-coef-eqn-MD1}) -- (\ref{eqn:adj-coef-eqn-MInvGauss1}), except that in this case we change $\bar{X}$ to $1/\lambda$. We see that the trends are similar to the cases with Poisson arrival. The upper bound given by (\ref{eqn:outage-inequality}) is quite tight.  We again see that the deterministic arrival gives the best performance, while the chi-squared arrival performs the worst. Comparing Fig.  \ref{fig:eventual-outage-for-exponential} for $\rho = 1.1$ with this figure, we see that for exponential inter-arrival times (Poisson arrival) the plot is close to that of the inverse Gaussian case. Also, the similarity between this figure and Fig. \ref{fig:eventual-outage-comparison-poisson-arrival} suggests a further approximation. For the sake of approximation, we have also plotted the eventual outage probability using the computed value of $r^*$ from (\ref{eqn:adj-coef-eqn-GM1}) in (\ref{eqn:eventual-outage-poisson-exact}) for the case of Poisson arrival. The closeness of the Monte Carlo plots to the lines given by  (\ref{eqn:eventual-outage-poisson-exact}) suggests that equation (\ref{eqn:eventual-outage-poisson-exact}) can serve as a reasonable approximation to the eventual outage probability when the energy arrival is given by a more general renewal process. That is, $\psi(u_0) \approx \psi_{\text{Poi}}(u_0)$.  
  
Fig. \ref{fig:battery-energy-cdf} shows the distribution of the battery energy when the self-sustainability condition is violated (i.e. when $\rho < 1$). Under this condition, the battery energy process achieves a steady state (ergodic and stationary) behavior; and it becomes meaningful to talk about the outage probability, $P_{out}$. We compare the empirical distribution of the battery energy with that obtained from heavy traffic approximation given in Section~VII.B (Discussion~6). For the empirical distribution, we assume Poisson arrival with exponentially distributed size of the energy packet. We see that the heavy traffic approximation fits reasonably with the empirical distribution when the battery energy level is high or when the value of $\rho$ is close to 1. We can also see that the outage probability $P_{out} = F_W(0)$ as obtained from empirical distribution fits well with the predicted values of $1-\rho$ as given in \tbf{Proposition \ref{prop:outage-prob-formula}}.


\section{Applications and Future Work}
\subsection{An Application to Wireless Communications}
A simple application of the concept of self-sustainability in the context of communication is as follows. In our basic model (\ref{eqn:energy-surplus}), let us assume that the incoming energy $X_i$ also encodes information via some form of amplitude modulation. Let the consumer be a receiver circuit with circuit power $p$. Thus we have a \tit{simultaneous wireless  information and power transfer} (SWIPT) system. Also, let the inter-arrival time $A$ between two energy packets be deterministic, so that $A = T$. Then, we have from the self-sustainability criteria of the energy harvesting system that $\frac{1}{T} > \frac{p}{\bar{X}}$.
From the point of view of information transfer, if $B$ is the bandwidth of the information channel, then by Nyquist ISI criterion, in order to avoid inter-symbol interference, the frequency of channel use should be $\frac{1}{T} > 2B$. Combining these two inequalities gives
\[ \frac{1}{T} > \max \left(2B, \frac{p}{\bar{X}} \right). \]
Lastly, let there be $Q$ possible sizes of energy packets, each size representing an information symbol. Thus, transmitting a single energy packet would represent $\log_2 Q$ bits of information and the data rate of the system would be $R_b = \frac{\log_2 Q}{T}$. Hence, for the SWIPT system to be self-sustainable as well as to  avoid inter-symbol interference, the minimum data rate should be 
\[ R_b  > \max \left(2B, \frac{p}{\bar{X}} \right) \log_2 Q. \]
This, however, does not guarantee error free communication if the $R_b$ is greater than the capacity of the information channel. 

\subsection{On the Performance Metrics of the Consumer}
Once energy has been harvested into the battery, the consumer puts the harvested energy into some use. Very often, the utility that the consumer derives by consuming the energy is a function of power, $p$. As such, for any utility which is a function of the power consumed, say $f(p)$, the consumer obtains $f(p)$ utility when the battery is not empty, $W>0$, and $f(0)$ when the battery is empty, $W=0$. Thus, we have the expected utility given by
\[ \Ebb[f(p)] = f(p)P(W>0) + f(0) P(W=0), \]
where the expectation is taken over the battery energy $W$. 

As noted in \tbf{Section \ref{sec:battery-energy-evolution}}, when $\rho >1$,  the energy outage probability $P_{out} = P(W=0) = 0$ for fixed power $p$. Therefore, we have two distinct behavior of the system, depending on the value of $\rho$:
\[ \Ebb[f(p)] = \left\{ \begin{array}{lr} 
				f(p) & \rho > 1, \\
				 f(p)P(W>0) + f(0) P(W=0) & \rho <1.
			     \end{array} \right. \]
It is interesting to note that the consumer is totally decoupled from the randomness of energy harvesting when $\rho > 1$. When $\rho < 1$, the $W$ becomes stationary and ergodic, and $P(W > 0) = \rho$ as implied by \tbf{Proposition \ref{prop:outage-prob-formula}}. Thus, the performance is limited by energy outage.

In green communication systems, where the harvested energy is used by the transmitter to transmit information, common utility functions like signal-to-interference-plus-noise-ratio (SINR) and throughput are zero when the transmit power $p=0$. That is, for these utility functions $f(0) = 0$. Other utility functions like SINR outage probability have $f(0)=1$. 

For instance, when throughput, $C(p)$, is considered as the utility function, since $P(W>0) = \rho$, we have
\[ \Ebb[C(p)] = \left\{ \begin{array}{lr} 
				C(p) & \rho > 1, \\
				 \rho \, C(p)   & \rho <1.
			     \end{array} \right. \]
where $C(0) = 0$. 

Likewise when the SINR outage probability, $P(O) = P(SINR< \theta)$, where $\theta$ is the threshold SINR, is taken as the utility function, we have by total probability theorem
\[ P(O) = P(O | W>0) P(W >0) + P(O | W =0)P(W=0). \]
Since the transmitter cannot transmit any information when the battery is empty, we have $P(O|W=0) = 1$. Thus,  
\[ P(O) = P(O|W=0)P(W>0) + P(W=0). \]
When $\rho > 1$, we know that $P(W=0)=0$ and $P(W > 0) = 1$. Therefore, for this case
\[ P(O) = P(O|W>0) \quad \mathrm{if} \quad \rho >1 \]
However, when $\rho <1$, we have $P(W = 0) = 1 - \rho$. Thus,
\begin{align*} 
P(O) &= \rho P(O|W>0) + 1-\rho \\
&= 1 - \rho (1 - P(O|W>0)).
\end{align*}
Putting everything together, we have
\[ P(O) = \left\{ \begin{array}{lr} 
				P(O|W>0) & \rho > 1, \\
				1 - \rho (1 - P(O|W>0)) & \rho <1.
			     \end{array} \right. \]

\subsection{Future Work}
We have defined the self-sustainability of an energy harvesting system and examined the case with some restrictive assumptions. We can change the basic assumptions and investigate the effects of these changes on the self-sustainability of an energy harvesting system. These will require new analysis. The following are a few open queries:
\begin{enumerate}
\item What becomes of the concept of self-sustainability when the battery capacity is finite?
\item How do we deal with periodic variations in harvest and consumption? 
\item How do we deal with consumption that varies with the battery state? 
\item How do we deal with the case when both consumption and harvest are stochastic?
\item What is the optimal consumption strategy when we consider self-sustainability as the performance criteria?
\item How should we study finite time horizon problems?
\item What should we do when the MGFs of the distributions do not exist?
\item Since the analogy between energy harvesting system and a queuing system is correct, is it possible to study a network of energy harvesting systems?
\item In the context of energy harvesting communications, what is the connection between the channel capacity and the self-sustainability probability?  We conjecture that capacity achieving strategies in green communication system are all self-sustaining, and vice versa. 
\end{enumerate}


\section{Conclusion}
We have given a mathematical definition of the concept of self-sustainability of an energy harvesting system, based on the concept of eventual energy outage. We have analyzed the harvest-store-consume system with infinite battery capacity, stochastic energy arrivals, and fixed energy consumption rate. The necessary condition for self-sustainability has been identified, and general formulas have been given relating the eventual outage probability and self-sustainability probability to various aspects of the underlying random walk process. Due to the complexity of the resulting formulas, assuming the existence of an adjustment coefficient, an exponential upper bound as well as an asymptotic formula has been obtained. Exact formulas for Poisson arrival process was also obtained. Lastly, the harvest-store-consume system has been shown to be equivalent to a $GI/G/1$ queueing system. Using the queueing analogy, outage probability can be easily found, in case the self-sustainability condition is not satisfied. Numerical results have been given to verify the analytical results.


\bibliographystyle{IEEE}

\end{document}